\UseRawInputEncoding
\documentclass[reprint,superscriptaddress,amsmath,amssymb,amsfonts,aps,prl,floatfix,twocolumn]{revtex4-2}
\usepackage{times}
\usepackage{graphicx}
\usepackage{dcolumn}
\usepackage{bm}
\usepackage{ulem}
\usepackage[colorlinks=true, citecolor=blue, linkcolor=blue, urlcolor=blue]{hyperref}
\usepackage{bbold}
\DeclareMathAlphabet\mathbfcal{OMS}{cmsy}{b}{n}
\usepackage{wasysym}
\usepackage{amssymb}
\usepackage{amsmath}
\usepackage[usenames,dvipsnames]{xcolor}
\usepackage{tikz}

\begin{document}

\title{Quantum Geometry and Stabilization of Fractional Chern Insulators Far from the Ideal Limit}
\author{Gal Shavit}
\affiliation{Department of Physics and Institute for Quantum Information and Matter, California Institute of Technology,
Pasadena, California 91125, USA}
\affiliation{Walter Burke Institute of Theoretical Physics, California Institute of Technology, Pasadena, California 91125, USA}
\author{Yuval Oreg}
\affiliation{Department of Condensed Matter Physics,    Weizmann Institute of Science, Rehovot, Israel 7610001}

\begin{abstract}
In the presence of strong electronic interactions, a partially filled Chern band may stabilize a fractional Chern insulator (FCI) state, the zero-field analog of the fractional quantum Hall phase.
While FCIs have long been hypothesized, feasible solid-state realizations only recently emerged, largely due to the rise of moir\'e materials.
In these systems, the quantum geometry of the electronic bands plays a critical role in stabilizing the FCI in the presence of competing correlated phases.
In the limit of ``ideal'' quantum geometry, where the quantum geometry is identical to that of Landau levels, this role is well understood.
However, in more realistic scenarios only empiric numerical evidence exists, accentuating the need for a clear understanding of the mechanism by which the FCI deteriorates moving further away from these ideal conditions.
We introduce and analyze an anisotropic model of a $\left|C \right|=1$ Chern insulator, whereupon partial filling of its bands, an FCI phase is stabilized over a certain parameter regime. 
We incorporate strong electronic interaction analytically by employing a coupled-wires approach, studying the FCI stability and its relation to the the quantum metric.
We identify an unusual anti-FCI phase benefiting from non-ideal geometry, generically subdominant to the FCI.
However, its presence hinders the formation of FCI in favor of other competitive phases at fractional fillings, such as the charge density wave.
Though quite peculiar, this anti-FCI phase may have already been observed in experiments at high magnetic fields.
This establish a direct link between quantum geometry and FCI stability in a tractable model far from any ideal band conditions, and illuminates a unique mechanism of FCI deterioration.
\end{abstract}

\maketitle

\textit{Introduction.---}
The fractional Chern insulator (FCI)~\cite{FCIbernevigPRX,FCIreviewNeupert_2015} is the lattice analog of the fractional quantum hall phase (FQH)~\cite{IntroFQHexp2,IntroFQLaughlin,IntroFQHaldane}, where strong correlations between electrons give rise to an extraordinary quantum phase of matter hosting exotic anyonic excitations ~\cite{FQHnobel,FQHhalperinjain}.
Unlike the FQH, the FCI may arise even in the absence of a magnetic field~\cite{FCIneuportPRL,Jackson2015Roynumerical,BergholtzLatticeConstantBerry}.
Recently, FCI phases were observed in moir\'e graphene devices~\cite{GraphenFCIYoung,FCIyacoby}, moir\'e transition metal dichalcogenides~\cite{FQAHseattle,FCIcornell,FCItransportSeattle,FCItransport_Shanghai}, and crystalline graphene multilayers~\cite{pentalayer_lu2023fractional}.

FCIs emerge out of a topologically-non-trivial band, whose dispersion is flat enough, such that correlations may stabilize the fractional phase.
However, these conditions are apparently  insufficient to guarantee FCI formation~\cite{RoytraceconditionFCI,BergholtzLatticeConstantBerry,FCI_Khalef_moire,FCI_TBG_parker2021fieldtuned}.
The quantum geometrical properties of the band have been argued to play a pivotal role in that regard.
Namely, bands whose geometries exactly mimic that of the lowest Landau level (LLL) exhibit an exact FCI ground-state under certain conditions~\cite{FQH_Exact_Kivelson,FQH_Exact_Haldane,FCI_exact_Bo,ledwith2022vortexability}.

Away from this exact "ideal" limit, several quantum geometry indicators have been proposed as a ruler to quantify how non-ideal the band is with respect to LLL~~\cite{simonrudnerindicators}.
These are substantiated by numerical evidence, supporting their relation to FCI stability~\cite{Jackson2015Roynumerical,BergholtzLatticeConstantBerry,FCI_TBG_parker2021fieldtuned,numerical_tmd_pressure_PhysRevResearch.5.L032022,TMD_numerical_Devakul,TMD_numerical_wang2023fractional}.
However, to date, there is no clear understanding of the relation between geometry indicators and the deterioration of the FCI phase in a strongly correlated band, especially in more realistic scenarios and far from the ideal limit.

In this Letter, we establish a direct link between FCI stability, electron-electron interaction parameters, and quantum geometrical properties of the strongly-correlated band hosting it.
We introduce a special coupled wires construction~(CWC)~\cite{wireconstructionKanePrl,wireconstructionkaneteoPrb,SagiFCIconstruction}, which we utilize to study fractional fillings of a Chern band.
The CWC employed allows one to study the competition between the FCI and competing phases, e.g., charge density wave~(CDW), as a function of tunable quantum geometry.
We exploit the inherent anisotropy of the model to gain an understanding of the effect of electron-electron interactions by employing bosonization techniques.
Crucially, in the presence of such interactions, non-ideal geometry promotes an anomalous phase, the aFCI, that impedes the FCI and may be experimentally revealed at high magnetic fields.
To characterize the suppression of FCI by this competition, we introduce a length scale that is directly related to relevant quantum geometry indicators.
Our tractable model thus illuminates the connection between quantum geometry,the strength of electron-electron interaction, and the emergence of FCIs away from ideal conditions normally considered.


\textit{Practical Chern insulator CWC.---}
We begin by considering an array of identical one-dimensional wires hosting spinless non-interacting fermions.
The interwire distance is $d$, and the intrawire unit-cell size is $2a$.
There are two states per unit cell (which allows for non-trivial topology), and we define the filling factor $\nu=2adn$, with $n$ the density

\begin{figure}
    \includegraphics[width=8.6cm]{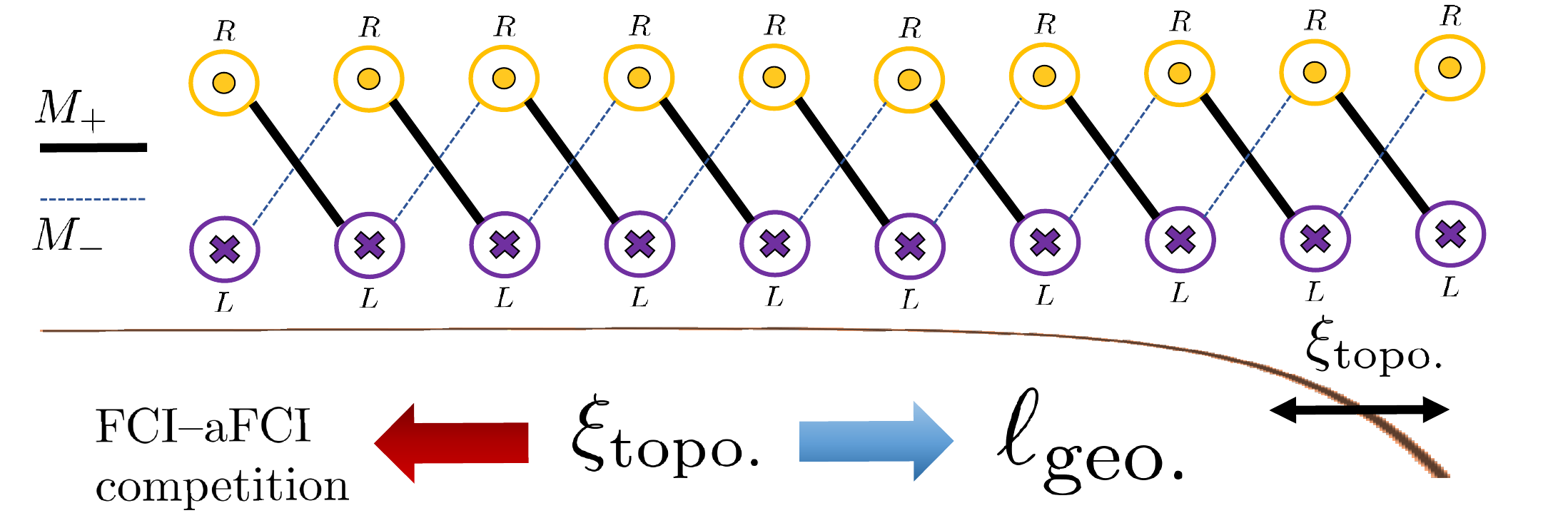}
    \centering
    \caption{Schematic depiction of the CWC, Eq.~\eqref{eq:HoCIcwc}.
    $R/L$ denote right/left moving fermions in each wire (going into or out of the page). 
    The couplings $M_+$ and $M_-$ are denoted by bold black and dashed blue lines, respectively.
    The extent of the edge-mode is $\xi_{\rm topo.}$, Eq.~\eqref{eq:xitopo}.
    In this work we demonstrate the analytic connection between quantum geometry, characterized by the length scale $\ell_{\rm geo.}$ [Eq.~\eqref{eq:geometriclengthMAIN}] and FCI stability.
    This is established through $\xi_{\rm topo.}$, which controls both the relative strength of the anti-FCI phase, and the band quantum geometry.
    }
\label{fig:schmeaticfigure}
\end{figure}

We consider the Hamiltonian (see Fig.~\ref{fig:schmeaticfigure} )
\begin{equation}
    H_0=\int dx\sum_{jj'}\Psi_{j}^{\dagger}\begin{pmatrix}\hat{\epsilon}_{F} & M_{-}\delta_{j,j'+1}\\
M_{+}\delta_{j,j'+1} & \hat{\epsilon}_{F}^*
\end{pmatrix}\Psi_{j'}+{\rm h.c.},\label{eq:HoCIcwc}
\end{equation}
where $\Psi_j=\left(\psi_{j,R},\psi_{j,L}\right)^T$ is a spinor of right/left moving ($R/L$) fermionic annihilation operators at position $x$ in wire $j$,
$\hat{\epsilon}_{F}=\frac{v_{F}}{2}\left(i\partial_{x}+k_{F}\right)\delta_{jj'}$, and $k_{F}=\frac{\pi}{2a}\left(1-\nu\right)$.
The $M_\pm$ terms couple opposite-chirality fermions on neighboring wires.
Time-reversal symmetry is broken whenever $\left|M_+\right|\neq \left|M_-\right|$, opening a gap at half-filling $\nu=1$, $E_{\rm gap}=2\left| \left|M_+\right| - \left|M_-\right|\right|$. 
The resultant bands have a Chern number $\left|C\right|=1$, where, e.g., the valence band has $C=1$ if $\left|M_+\right|>\left|M_-\right|$ (which we will assume henceforth without loss of generality), and vice-versa. 
Eq.~\eqref{eq:HoCIcwc} may be obtained as the low-energy description of a lattice model with zero magnetic flux per unit-cell, which maps to an anisotropic version of the half-Bernevig-Hughes-Zhang (BHZ) model~\cite{BHZmodelScience},  see the text and Fig.~\ref{figapp:2dlatticeschematics} in the Supplementary Materials (SM)~\cite{SM}.

A length scale that will prove crucial to our discussion is the transverse-direction extent of the topological chiral edge-states~\cite{sshcorrelationlength},
\begin{equation}
    \xi_{\rm topo.}^{-1} = \frac{1}{2}\log\frac{\left|M_+\right|}{\left|M_-\right|}.\label{eq:xitopo}
\end{equation}
Notice that if one mass term vanishes, $H_0$ is equivalent to the well-known CWC of the lowest Landau level~(LLL)~\cite{wireconstructionKanePrl,wireconstructionkaneteoPrb}, $\xi_{\rm topo.}\to 0$, and the edge-states confine to a single wire.
We will refer to this as the \textit{optimal} CWC. 
We note that although $\xi_{\rm topo.}$ represents the edge extent of the chiral edge mode, it is actually a bulk property that does not depend on the boundary condition. Its divergence indicates a transition between $C=1$ and $C=-1$.

We now turn to discuss the quantum geometry properties of the bands of $H_0$, captured by the momentum space tensor
\begin{equation}
    \eta_{\alpha\beta}\left({\mathbf k}\right)=
    \left\langle \partial_\alpha u_\mathbf{k}| \partial_\beta u_\mathbf{k} \right\rangle
    - 
    \left\langle \partial_\alpha u_\mathbf{k}\left|u_{\mathbf{k}}\right\rangle\left\langle u_{\mathbf{k}}\right| \partial_\beta u_\mathbf{k} \right\rangle,
    \label{eq:quantumgeometricaltensorMAIN}
\end{equation}
where $|u_{\bf k}\rangle$ is the wavefunction of the valence band at momentum ${\bf k}=(k_x,k_y)$, and $\partial_\alpha = \frac{\partial}{\partial k_{\alpha}}$.
The Berry curvature is $\Omega  = 2{\rm Im}\eta_{yx}$, and the Fubini-Study metric is given by $g_{\alpha\beta}={\rm Re}\eta_{\alpha\beta}$.
These satisfy the inequality ${\rm tr}\, g \geq \left|\Omega\right|$.
It has been shown~\cite{RoytraceconditionFCI,ledwith2022vortexability} that for a band with flat Berry curvature which
saturates this inequality, the density operators projected onto that band reproduce the Girvin-MacDonald-Platzmann (GMP) algebra~\cite{GMP_PhysRevB.33.2481} of the~LLL. 

As such, the deviation from the so-called ``trace condition'' may be quantified by
$\Bar{T} \equiv \int \frac{d^2{\mathbf k}}{A}\left({\rm tr} g - \left|\Omega\right|\right)$, where $A$ is the area of the Brillouin zone (BZ).
The BZ integral over ${\rm tr}\,g$ has been shown to correspond to the minimal Wannier-function spread associated with a set of bands~\cite{Vanderbiltmaximmalylocalized}.
Similarly, let us examine the length scale
\begin{equation}
    \ell_{\rm geo.}=4 \int\frac{dk_y}{2\pi}{\rm tr} g \left(k_x=0,k_y\right),\label{eq:geometriclengthMAIN}
\end{equation}
which, in our model, constitutes a major contribution to $\Bar{T}$, and the most important one for our purposes~\footnote{We have verified that the length scale $\ell_{\rm geo.}$ is closely correlated to the trace condition violation $\Bar{T}$ in a parent lattice model, justifying our focused attention on the former as an extension of the latter, see SM~\cite{SM}.}.
We calculate $\ell_{\rm geo.}$ in terms of $d/\xi_{\rm topo.}$ and $\alpha \equiv v_F/\left(dM_+\right)$~\cite{SM}.
Close to the optimal CWC, $\xi_{\rm topo.}\ll d$, $\ell_{\rm geo.}$ is minimal and is approximately $d\left(1+\alpha^2\right)$.
In the opposite limit, we find 
$\ell_{\rm geo.} \left(\xi_{\rm topo.}\gg d\right) \approx \xi_{\rm topo.}\left(1+\alpha^2/4\right)$.

Relating the quantum geometry of the band to the correlation length $\xi_{\rm topo.}$ is one of our key results.
This establishes the adverse effect of having both types of chiral coupling $M_\pm$ in the CWC on ${\rm tr}\,g$, in the sense of pushing it further away from its lower bound, and rendering the trace-condition-violation $\Bar{T}$ larger.
We have also verified that FCI indicators for the lattice model related to $H_0$ are optimized close to when one of the mass term dominates, i.e., when $\xi_{\rm topo.}$ is small (see SM~\cite{SM}).
Next, we will demonstrate how large $\xi_{\rm topo.}$ potentially leads to the destabilization of FCI.

\textit{Fractional filling and interactions.---}
The wire construction presented in this Letter has a periodic modulation along the wires. This allows for the stabilization of FCI by adjusting the electron density~\footnote{The construction presented here differs significantly from that of Ref.~\cite{SagiFCIconstruction}, where the amplitude of an alternating magnetic field determines the effective filling factor.}
For concreteness, we focus on Laughlin-like fractional filling of the valence band, $\nu_{\rm frac.}=\left(2p+1\right)^{-1}\equiv 1/m$, with $p$ a positive integer.

To account for interactions, we employ the framework of abelian bosonization~\cite{giamarchi2004quantum,1dFL,LLHaldane}.
We represent the chiral fermionic operators in terms of bosonic variables,
$\psi_{j,r}\sim \frac{1}{\sqrt{2\pi a}}
e^{irk_F-i\left(r\phi_j-\theta_j\right)}$, with $r=1$ (-1) for right (left) movers, and the algebra
$\left[\phi_j\left(x\right),\partial_x \theta_j'\left(x'\right)\right]
=i\pi\delta_{jj'}\delta\left(x-x'\right)$
is satisfied.

The bosonic version of $H_0$, supplemented by forward-scattering interactions may be written as,
\begin{equation}
    H_{\rm f.s.} = \int dx \sum_{jj'}
    \partial_x\chi_j
    {\cal M}^{jj'}
    \partial_x\chi_{j'}^T,
\end{equation}
with $\chi_j = \left(\phi_j,\theta_j\right)^T$, and
${\cal M}^{jj'}=\frac{v_F}{2\pi}I\delta^{jj'}+{\cal U}^{jj'}$, where $I$ is the unit $2\times2$ matrix, and $\cal U$ describes the interactions.
Notice that the single-particle terms $M_\pm$ do not conserve momentum away from $\nu=1$, and are thus absent from this low-energy description.
Their important role, however, will be clarified shortly.

We now include large-momentum scattering interactions, comprising processes with several of the operators
${\cal O}_{j,{\rm bs}}=\psi_{j,R}^\dagger \psi_{j,L}$.
A Laughlin-like FCI phase, with Chern number $C=1/m$ may be stabilized by the operator
\begin{equation}
    {\cal O}^j_{\rm FCI} \sim g_{\rm FCI} 
    \left({\cal O}_{j,{\rm bs}}\right)^p
    \left({\cal O}_{j+1,{\rm bs}}\right)^p
    \psi_{j,R}^\dagger \psi_{j+1,L}
    +{\rm h.c.} \label{eq:OFCImain}
\end{equation}
We note that at the filling $\nu_{\rm frac.}$, this term conserves momentum modulo-$\pi/a$.
In contrast to conventional CWC, momentum conservation is enabled by the underlying lattice, not by the external magnetic field (cf. Ref.~\cite{Loss_moire_cwc}).
Interestingly, this means that the time-reversal partner of ${\cal O}^j_{\rm FCI}$,
\begin{equation}
    {\cal O}^j_{\rm aFCI} \sim g_{\rm aFCI} 
    \left({\cal O}_{j,{\rm bs}}\right)^p
    \left({\cal O}_{j+1,{\rm bs}}\right)^p
    \psi_{j+1,R}^\dagger \psi_{j,L}
    +{\rm h.c.} \label{eq:OaFCImain}
\end{equation}
may be stabilized for the same reason.
The stabilized phase, however, has $C=-1/m$, so we refer to it as the anti-FCI (aFCI).
Such a term is forbidden in LLL CWCs by momentum conservation, and its appearance is \textit{unique} to our proposed framework.
Crucially, however, it is clear that the FCI and aFCI terms are not on equal footing.
Both include an interwire part, yet $g_{\rm FCI}\propto M_+$, and $g_{\rm aFCI}\propto M_-$.
Thus,assuming $M_+>M_- $, the FCI should always prevail over the aFCI by construction.
Nevertheless, we will demonstrate that the aFCI destabilizes the FCI phase, thus relating FCI stability to small $\ell_{\rm geo.}$ and favorable quantum geometry. 

We consider an additional multi-particle process at this filling, stabilizing a CDW,
\begin{equation}
    {\cal O}^j_{\rm CDW} = g_{\rm CDW}\left({\cal O}_{j,{\rm bs}}\right)^{2p+1}
    +{\rm h.c.}
\end{equation}
This term too is enabled by the presence of the lattice, and is absent from quantum-Hall CWCs.
The CDW coincides with the FCI in the thin-torus limit~\cite{TaoThouless}, and is distinct from the bubble/stripe phases potentially stabilized at high Landau levels~\cite{FoglerCDWPRL1996,FoglerCDWPRB1996}, not considered in our analysis~\footnote{In our model, the corresponding phase with domains of alternating densities would be a stripe phase. Within such a phase, the densities at different wires would fluctuate significantly, a scenario which is not captured within our Luttinger liquid description. We leave the study of the role of the stripe phase in Chern bands with poor quantum geometry for future work.}.
To see that it stabilizes a CDW, consider its bosonized form $\propto g_{\rm CDW}\cos\left[2\left(2p+1\right)\phi_j\right]$.
At strong coupling, $\phi_j\left(x\right)$ sets at a uniform value $\phi_0$, the density operator is modulated periodically along the wires and can be approximated by
$\rho\left(x\right)\propto\cos^2\left(\pi\nu_{\rm frac.}x - \phi_0\right)$~\cite{giamarchi2004quantum}.
The relative phase between CDWs in neighboring wires is determined by including the interaction terms
${\cal O}^{j}_{\phi}\sim g_\phi {\cal O}^\dagger_{j,{\rm bs}}{\cal O}_{j+1,{\rm bs}}+{\rm h.c.}$ (these conserve momentum regardless of density).

\textit{Weak-coupling RG.---}
The competition between the FCI, aFCI, and CDW phases can be readily understood by considering the renormalization group (RG) flow of the low-energy theory.
Considering weak-coupling, the most salient conclusions can be derived from a simplified two-wire model.
The non-commutativity between, e.g., ${\cal O}_{{\rm FCI}}^{j}$ and ${\cal O}_{{\rm aFCI}}^{j+1}$, would manifest in higher orders in the RG flow equations, motivating the two-wire approach.
In the SM, we discuss the strong-coupling limit and argue that within that limit, the FCI phase has a many-body gap proportional to the difference~$g_{\rm FCI}-g_{\rm aFCI}$~\cite{SM}.

The simplified Hamiltonian is
\begin{align}
    {\cal H}	&=\sum_{i=\rho,\sigma}\frac{u_{i}}{2\pi}\left[K_{i}^{-1}\left(\partial_{x}\phi_{i}\right)^{2}+K_{i}\left(\partial_{x}\theta_{i}\right)^{2}\right]\nonumber\\
	&+\frac{\tilde{V}}{2\pi}\left(\partial_{x}\phi_{\rho}\partial_{x}\theta_{\sigma}+\partial_{x}\theta_{\rho}\partial_{x}\phi_{\sigma}\right)\nonumber\\
	&+\sum_{j}\left({\cal O}_{{\rm FCI}}^{j}+{\cal O}_{{\rm aFCI}}^{j}+{\cal O}_{{\rm CDW}}^{j}+{\cal O}_{{\rm \phi}}^{j}+{\rm h.c.}\right).
\end{align}
The two sectors $\rho,\sigma$ correspond to combinations of the fields on the two wires labeled $1,2$, e.g., $\phi_{\rho/\sigma}=1/\sqrt{2}\left(\phi_1\pm\phi_2\right)$.
The $\tilde{V}$ term breaks time-reversal symmetry, and is generated by the RG flow of the FCI/aFCI terms.
The FCI terms have the same scaling dimension $d_{\rm FCI}=\frac{m^2}{2}K_\rho+\frac{1}{2}K_\sigma^{-1}$, whereas the CDW term has $ d_{\rm CDW} = \frac{m^2}{2}K_\rho+\frac{m^2}{2}K_\sigma$.
Clearly, $K_\rho$, which is expected to be rather small in the case of strong repulsive interactions, will not play any meaningful role in the competition between backscattering terms.
However, $K_{\rho}$ controls the transition between a gapped phase and a gapless sliding Luttinger liquid~\cite{sll_Ashvin,sll_Kane}.
Conversely, $K_{\sigma}$ directly relates to the competition between the CDW phase requiring sufficiently small $K_\sigma$, and the FCI terms which favor large values of $K_\sigma$.

We derive the RG equations using the operator product expansion (OPE)~\cite{cardy_1996}.
The short-distance cutoff is parameterized $\alpha=\alpha_0e^\ell$, where in each RG step, $\ell$ increases incrementally.
We define dimensionless coupling constants $y_i\equiv g_i / \left(\pi u\right)$, allowing us to obtain the RG flow equations, presented in full in the SM~\cite{SM}.

\begin{figure}
    \includegraphics[width=8.5cm]{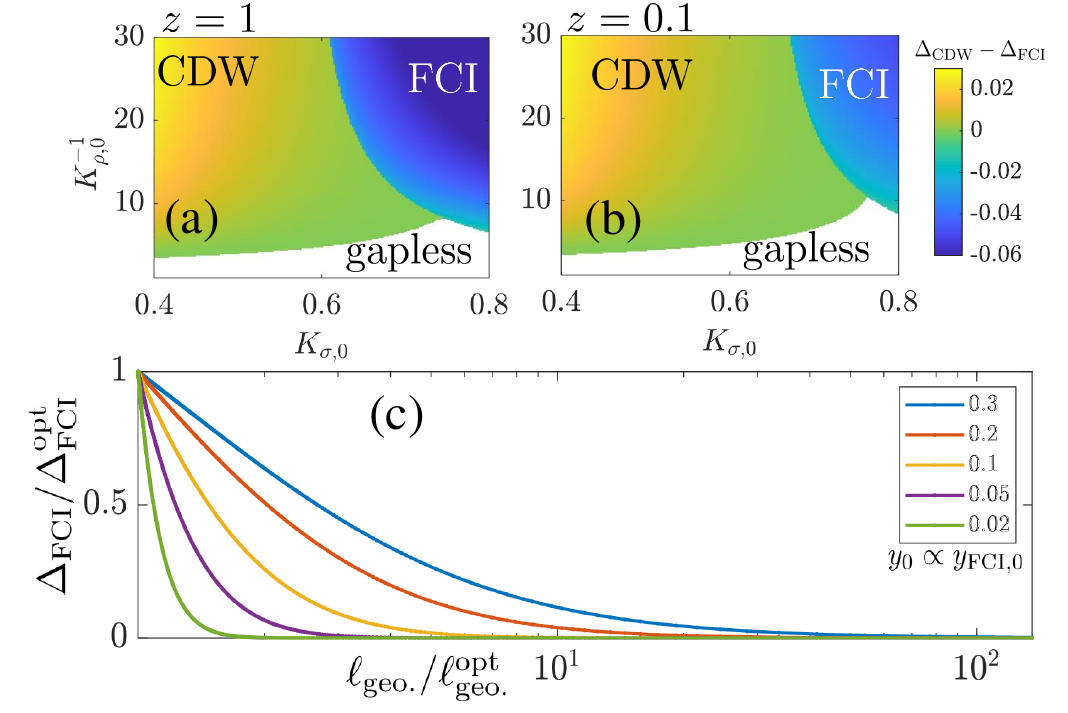}
    \centering
    \caption{(a)--(b)
    Phase diagram obtained by the RG flow equations~\cite{SM}.
    We plot the gap proxies, assigning opposite signs to CDW/FCI instabilities,
    $\Delta_{\rm CDW}-\Delta_{\rm FCI}$ (see text) for (a) optimal quantum geometry ($z=\tanh\frac{2d}{\xi_{\rm topo.}}=1$), and (b) ``poor'' geometry ($z=0.1$).
    Here $m=3$ and initial conditions are $y_{F,0}=0.03$, $y_{\rm CDW,0}=0.08$, $y_{\phi,0}=0.1$, $\tilde{V}_0=0$.
    (c)
    Effect of quantum geometry on the FCI gap proxy in the maximal-competition regime, $d_{\rm FCI}=2$.
    different values of $y_0=m\sqrt{\frac{K_\rho}{2K_\sigma}}y_{\rm FCI,0}$ are indicated by the legend.}
\label{fig:mainfigure}
\end{figure}

In Fig.~\ref{fig:mainfigure}a,b we present examples of the phase diagram obtained from integration of the RG equations.
The RG flow is stopped when either $y_{\rm FCI/CDW}>1$, obtaining the strong-coupling scale $\ell_{{\rm FCI/CDW}}$.
The proxy for the relevant gap is evaluated as $\Delta_{i}\equiv\exp\left(-\ell_i\right)$.
We parameterize the initial conditions as $y_{F,0}^2=y_{\rm FCI,0}^2+y_{\rm aFCI,0}^2$, $zy_{F,0}^2=y_{\rm FCI,0}^2-y_{\rm aFCI,0}^2$, noting $z=\tanh\frac{2d}{\xi_{\rm topo.}}$.
Moving away from the optimal $z=1$ towards $z\ll 1$, the region in the phase diagram where FCI is stabilized dramatically shrinks, and its gap weakened.
Thus, quantum geometry directly impacts FCI stability through competition with the hidden aFCI phase

Crucial insight is gained by considering the scenario $d_{\rm FCI}\to 2$, where the aforementioned competition is most pronounced.
In this limit, one only considers the RG flow of $y_{\rm FCI/ aFCI}$ and $\tilde{V}$, which realize a generalized Berezinskii-Kosterlitz-Thouless (BKT) flow~\cite{SM}.
If initially $\tilde{V}=0$, we find a closed-form expression for the FCI-divergence RG time, 
$\ell^{\infty}=\frac{u}{m\sqrt{\frac{K_\rho}{2K_\sigma}}y_{{\rm FCI},0}}{\rm Re}\left[K\left(u^{2}\right)\right]$,
where $K\left(x\right)$ is the complete elliptic integral of the first kind, and 
$u=\sqrt{\frac{1+z}{1-z}}$. 
Fig.~\ref{fig:mainfigure}c plots the FCI gap proxy $e^{-\ell^\infty}$, showing the rapid FCI deterioration as the quantum geometry becomes far from optimal.
Away from optimal quantum geometry, $u\approx 1$, we can analytically relate the energy scale to the quantum metric,
\begin{equation}
    \Delta_{\rm FCI}\propto \ell_{\rm geo.}^{-\frac{\sqrt{K_{\sigma}/\left(2K_{\rho}\right)}}{my_{{\rm FCI},0}}},\label{eq:DeltaFCIanalytic}
\end{equation}
further stressing the connection between the FCI stability and quantum geometry via the FCI--aFCI competition.

\textit{External magnetic field.---}
A perpendicular magnetic field applied to the system has been argued to promote FCI formation through its impact on quantum geometry indicators.
Within our CWC model, however, we may directly probe its role.
We introduce the field $B$ through a boost $\psi_{j,r}\to\psi_{j,r}e^{ibjx}$, where $b=edB/\hbar$.
Whereas the CDW commensurability remains at $\nu_{\rm frac.}$, we find that the filling factor at which the fractional Chern processes conserve momentum are given by 
\begin{equation}
    \nu^*_{\rm FCI/aFCI} = \nu_{\rm frac.} \pm \frac{1}{m}{\Phi},
\end{equation}
where $\Phi$ is the number of $h/e$ flux-quanta per unit-cell.

\begin{figure}
    \includegraphics[width=8.7cm]{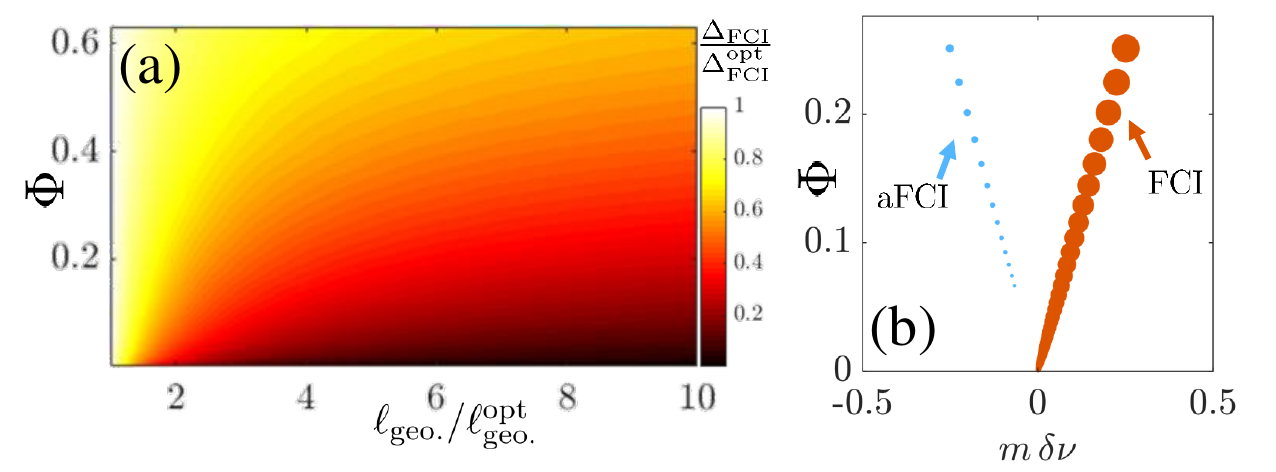}
    \centering
    \caption{(a)
    FCI gap proxy in the maximal competition regime along the FCI-commensurate line, where the deviation from fractional filling follows $\delta\nu=\Phi/m$.
    Higher flux relieves the FCI--aFCI competition, strengthening the gap, similar to the effect of improving the quantum geometry [$\ell_{\rm geo.}$ defined in Eq.~\eqref{eq:geometriclengthMAIN}].
    We use $y_0=m\sqrt{\frac{K_\rho}{2K_\sigma}}y_{\rm FCI,0}=0.05$.
    (b)
    Evolution of the gap proxies for FCI (orange) and aFCI (blue) phases, along the corresponding commensurate lines.
    The dots size is proportional to the gap.
    Here, $y_0=0.1$, and $\ell_{\rm geo.}=12\ell_{\rm geo.}^{\rm opt}$.}
\label{fig:magneticfigure}
\end{figure}

Thus, the magnetic field naturally \textit{relieves the FCI--aFCI tension}, as the two are stabilized at diverging densities.
The field also separates the CDW from the FCI, hence potentially favoring FCI formation even further.

We illustrate this effect by modification of the RG flow equations in the constrained $d_{\rm FCI}=2$ regime.
When the multiparticle backscattering terms are incommensurate, due to density deviations $\delta\nu=\nu-\nu_{\rm frac.}$ and/or finite $B$, they acquire a spatial oscillation period.
Within the RG flow, it is reasonable~\cite{commIncommRG} to treat this period as a soft cutoff on the effect of these terms.
Along the FCI-commensurate line $m\delta\nu=\Phi$, we impose this cutoff by multiplying $y_{\rm aFCI}^2$ in
the RG flow of $\Tilde{V}$
by~$c\left(\ell\right) = \left(1+e^{\gamma\left(\ell+\log{\Phi}\right)}\right)^{-1}$ ($\gamma$ controls the cutoff smoothness, set as $\gamma=2$ throughout our calculations).
As demonstrated in Fig.~\ref{fig:magneticfigure}a, the magnetic flux enhances the FCI stability, particularly so in the regions where quantum geometry is far from optimal.
We stress that this is entirely due to aFCI suppression, driven by its incommensurability with increased field.

A curious consequence of the FCI--aFCI diverging paths, is the possible emergence of the aFCI phase at high enough fields.
Following a similar cutoff treatment along the aFCI line $m\delta\nu=-\Phi$, we find this anomalous phase may indeed be stabilized at a different density, see Fig.~\ref{fig:magneticfigure}
b.
Surprisingly, Ref.~\cite{FCIyacoby} observed two high magnetic field features with fractional Chern numbers $-\frac{8}{5}$ and $-\frac{7}{3}$ moving towards lower densities as a function of magnetic fields~\cite{parker2021fieldtunedFCIledwitharxiv} (Fig. 1b in~\cite{FCIyacoby}, FCI features closest to $\nu=0$). 
This phenomenology is suggestive of an aFCI-like phase.

\textit{Conclusions.---}
We have studied analytically the connection between non-optimal (far from LLL-like) quantum geometry and the FCI stability in an anisotropic model of topological phases with correlated electrons.
The model we introduce features competition between a lattice-enabled CDW phases, a unique aFCI phase, and the FCI phase. 
This competition is unique to our model due to the periodic modulation of interwire hopping introduced along the wire.

The aFCI phase was demonstrated to be enhanced by non-optimal quantum geometry, subsequently critically suppressing the FCI phase.
Within this model, quantum geometry was shown to be intimately connected to the topological correlation length.
In turn, this length scale directly controls the relative strength of the interactions, which stabilize the adverse aFCI, enabling the CDW to take over as the leading instability.
We thus established an unambiguous connection between so-called FCI indicators far from non-realistic ideal scenarios and the potential stability of the FCI.
In a particular regime, an analytical expression relating the two has been presented, Eq.~\eqref{eq:DeltaFCIanalytic}.

Our model further illuminates the role of an external magnetic field.
Namely, the fractional fillings at which the competing correlated phases may be stabilized are ``pulled apart'' (Fig.~\ref{fig:magneticfigure}b) with increasing field, naturally suppressing the competition through induced incommensurability.
Possible signatures of the peculiar aFCI phase in high magnetic field were discussed, potentially already detected in experiments~\cite{FCIyacoby}.

These insights provided by our work into the stability of the exotic FCI under non-ideal conditions, as well as elucidation of the mechanism by which the FCI deteriorates in realistic bands, may be beneficial for material exploration and band engineering.
As an example, in the discussed model the terms ${\cal O}^j_\phi$ stabilize the CDW at the expanse of the FCI.
This role of longer-range interactions has been mentioned in numerical investigations~\cite{FCI_Fu_Crepel}.
Conversely, this interaction may be suppressed by imposing a periodic modulation of the density in the transverse direction~\cite{BLG_design_lattice_periodic,Periodic_Devakul_tan2024designing}, effectively dephasing the CDWs on neighboring wires~\cite{SM}.
Moreover, the model we propose can account for the curious phenomena of CDW stabilization by a magnetic field~\cite{SM}, recently observed in experiments~\cite{magneticWignerCrystal,magneticWignerCrystal2}.

\begin{acknowledgments}
This project was partially supported by grants from the ERC under the European Union’s Horizon 2020 research and innovation programme (grant agreement LEGOTOP No 788715), the DFG CRC SFB/TRR183, and the ISF Quantum Science and Technology (2074/19).
GS acknowledges support from the Walter Burke Institute for Theoretical Physics at Caltech, and from the Yad Hanadiv Foundation through the Rothschild fellowship.
\end{acknowledgments}

\bibliography{FCI.bib}

\pagebreak
\begin{widetext}

\global\long\def\thesection{S.\Alph{section}}%
 \setcounter{figure}{0} 
\global\long\def\thefigure{S\arabic{figure}}%
 \setcounter{equation}{0} 
\global\long\def\theequation{S\arabic{equation}}%
\section*{Supplemental Material for ``Quantum Geometry and Stabilization of Fractional Chern Insulators Far from the Ideal Limit''}

\setcounter{section}{0} \renewcommand{\thesection}{S.\arabic{section}} 
\setcounter{figure}{0} \renewcommand{\thefigure}{S\arabic{figure}} 
\setcounter{equation}{0} \renewcommand{\theequation}{S\arabic{equation}}

\section {Lattice model}\label{app:latticemodel}
Here, we describe a the two-dimensional tight-binding model which is the UV analog of the coupled-wire model described and studied in the main text.
The tight binding Hamiltonian is comprised of two terms,
\begin{equation}
    H_{\rm 2d} = H_{\rm wire} + H_{\rm interwire},\label{eq:tightbindingHamiltonian}
\end{equation}
where $H_{\rm wire}$ describes the physics along the effective wires direction, and $H_{\rm interwire}$ accounts for their coupling to each other.
We consider the simplest form possible for $H_{\rm wire}$, taking into account only nearest-neighbor hopping $t$,
\begin{equation}
    H_{\rm wire} = -t \sum_j \sum_m \left( c_{j,m+1}^\dagger c_{j,m} +{\rm h.c.}\right),
    \label{eq:Hwire}
\end{equation}
and $c_{j,m}$ annihilates a fermion on the $m$-th site of the $j$-th wire.
In order to synthesize a Chern insulator out of this plain array of quantum wires, we need some non-trivial form of the interwire hopping Hamiltonian $H_{\rm interwire}$.
It consists of two contributions with hopping amplitudes $t_1$ and $t_2$ (both real numbers) (see Fig.~\ref{figapp:2dlatticeschematics} for illustration),
\begin{align}
    H_{\rm interwire} &= 
     it_1\sum_j\sum_m \left(-1\right)^{m} c_{j+1,m}^\dagger c_{j,m} \nonumber\\
    &+
    t_2\sum_{j} \sum_m \sin^2\frac{\pi m}{2} \left(c_{j+1,m+1}^\dagger c_{j,m} + c_{j+1,m}^\dagger c_{j,m+1} \right)\nonumber\\
    & + \rm h.c.
    \label{eq:Hinterwire}
\end{align}
Notice that the inter-wire coupling reduces the translational symmetry to $m\to m+2$, i.e., doubles the unit-cell along the direction of the wires.
The $t_1$ term represents hopping of fermions with a $\pm \pi/2$, phase alternating between neighboring sites. 
This introduces a $\pi$-flux to each plaquette in the two-dimensional lattice.
This term is insufficient to induce a Chern insulating phase, as it conserves the systems compound time-reversal symmetry which combines complex conjugation $i\to -i$ and translation by half a unit-cell $m\to m+1$.
However, the $t_2$ coupling term explicitly breaks it.
Intuitively, it introduces possible hopping paths which encircle a time-reversal asymmetric flux (in contrast to the symmetric $\pi$-flux).

\begin{figure}
    \includegraphics[width=14cm]{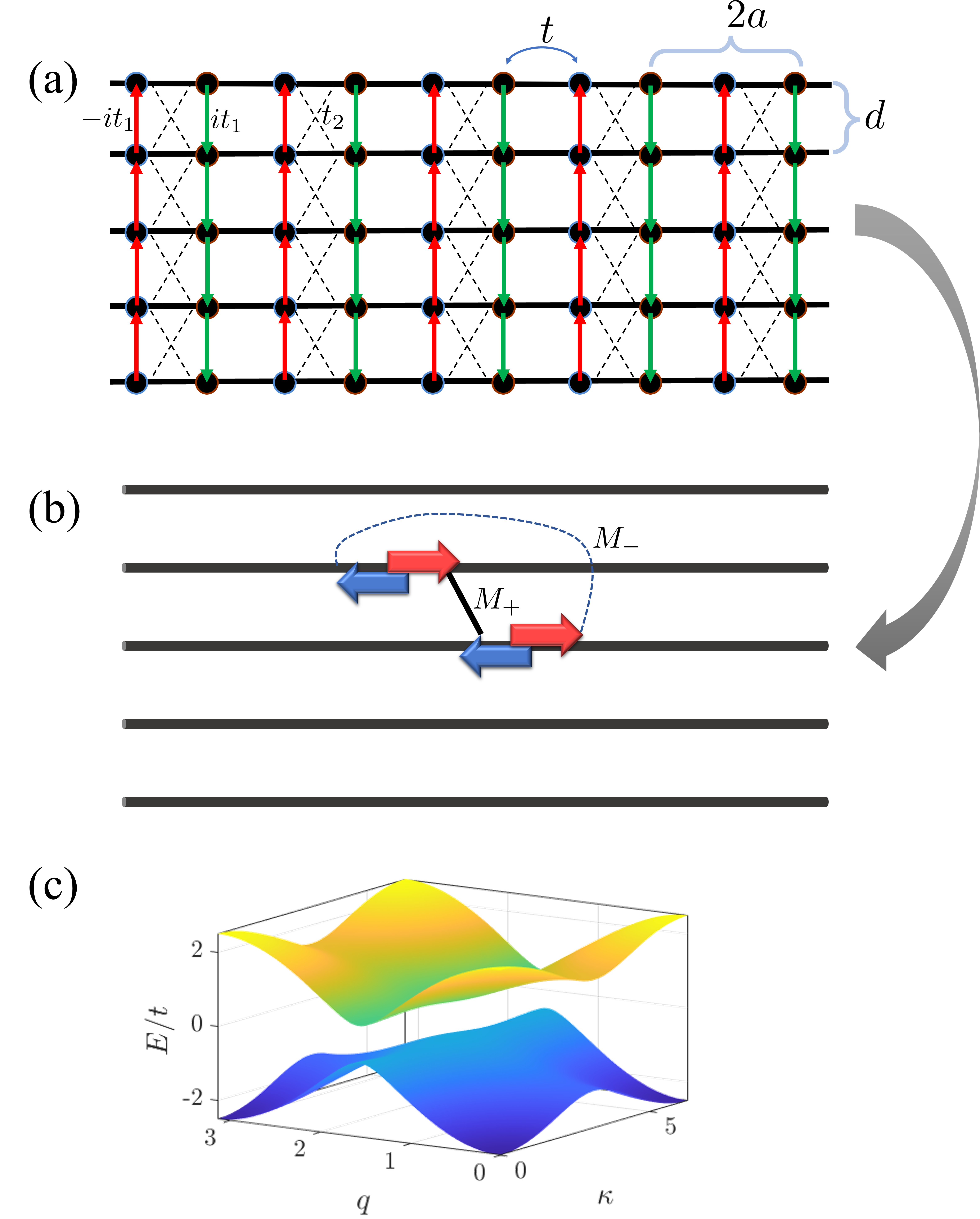}
    \centering
    \caption{
    (a)
    Schematic description of the two dimensional lattice model we discuss.
    Hopping along the horizontal direction (intra-wire) has the magnitude $t$ connecting nearest neighbors.
    An alternating sign imaginary coupling with amplitude $\pm i t_1$ (red and green arrows) connect two neighboring wires in the vertical direction.
    Finally, a real-valued hopping term cross couples opposite sublattice sites in neighboring wires with amplitude $t_2$ (dashed black lines).
    The interplay between the two inter-wire terms, $t_1$ and $t_2$, determines the topology of the resulting bands, as well as the corresponding quantum geometry.
    In the presence of the staggered hopping between the wires, the size of the intrawire unit-cell is $2a$, and the distance between the wires is $d$, as indicated.
    (b)
    The effective quasi-one-dimensional system, or coupled wires construction that is born of the lattice model, presented in Eq.~\eqref{eq:HoCIcwc} in the main text.
    Arrows in different direction correspond to opposite chirality low-lying modes, and their couplings by $M_+$ and $M_-$ are illustrated.
    (c)
    Band structure of the lattice model in momentum space [Eq.~\eqref{eq:HCIspectrum2d}], for $t_1=0.3t$, $t_2=0.2t$.
    It is clearly shown that the spectrum varies quite little along $\kappa$, indicative of weak coupling of the wires in the transverse direction.}
\label{figapp:2dlatticeschematics}
\end{figure}

\subsection{Ladder geometry and the continuum limit}
Before diagonalizing the Hamiltonian $H_{\rm 2d}$ and obtaining its spectrum, it is instructive to consider a ladder comprised of only two neighboring wires.
Making the unit-cell doubling incurred by the wire coupling explicit, we define the two species of fermionic annihilation operators in momentum space,
$A_{j,q}=L^{-1/2}\sum_{m\,{\rm even}}e^{imq}c_{j,m}$ 
and 
$B_{j,q}=L^{-1/2}\sum_{m\,{\rm odd}}e^{imq}c_{j,m}$.
Notice $q\in\left[0,\pi\right]$ defines the reduced Brillouin zone, and we set the lattice spacing to unity.
Introducing the spinor $\Psi_q=\left(A_{1,q},B_{1,q},A_{2,q},B_{2,q}\right)^T$, we write the two-wire Hamiltonian as
\begin{equation}
    H_{2-{\rm wire}}=\sum_{q}\Psi^\dagger_q
    \left[
    -t\left(1+\cos 2q\right)\sigma_x -t\sin 2q \sigma_y + t_1\sigma_z\tau_y + t_2\sigma_x\tau_x
    \right]
    \Psi_q,
    \label{eq:H2wirefull}
\end{equation}
where $\sigma_i$ and $\tau_i$ are Pauli matrices acting on sublattice and which-wire degrees of freedom, respectively.

Expanding the Hamiltonian to lowest order in $q-\pi/2$,  we find that when $t_1=t_2=0$, the low-energy spectrum of each wire is effectively described by chiral right/left moving annihilation operators $\psi_{j,R/L}$.
These can be understood in the sublattice language as the two eigenvectors of $\sigma_y$.

In the continuum limit, restoring all the wires in the system, the Hamiltonian effectively becomes
\begin{equation}
    H_{\rm eff.} =\sum_{j=1,2} \int dx \left( {\cal H}_D^j + {\cal H}_{M}^j\right), 
\end{equation}
\begin{equation}
    {\cal H}_D^j = v 
    \left(\psi_{j,R}^\dagger i\partial_x \psi_{j,R} - \psi_{j,L}^\dagger i\partial_x \psi_{j,L}\right),
    \label{eq:H2wiresDirac}
\end{equation}
\begin{equation}
    {\cal H}_{M}^j= M_+\psi_{j,R}^\dagger\psi_{j+1,L}+M_-\psi_{j,L}^\dagger\psi_{j+1,R}+{\rm h.c.}
\end{equation}
Here, the Fermi velocity of the chiral modes is related to the hopping strength $v=2t$, and the mass terms are $M_{\pm}=t_2\pm t_1$.
Thus, we recover precisely the form of the continuum coupled-wire approach we analyze thoroughly in the main text.
By tuning the microscopic interwire hopping $t_1$ and $t_2$, one may tune the strength of the chiral modes coupling between adjacent wires.

We note here that away from half filling ($q=\pi/2$) expansion of the two wire Hamiltonian reveals a single-particle hopping process which couples modes of the same chirality in neighboring wires, e.g., 
$t_\perp \psi_{j,R}^\dagger\psi_{j+1,R}$.
Its strength increases as one moves away from half-filling,
\begin{equation}
    t_\perp \approx -t_{2}\left|\cos q_{F}\right|. \label{eq:supptperp}
\end{equation}
In this work we consider strongly interacting systems, where the single particle hopping terms between the wires are highly irrelevant.
It will thus turn out that $t_{\perp}$ has little to no effect on our model, except for perturbatively enabling certain scattering channels in various scenarios, see Fig.~\ref{fig:perturbationcouplingconstants}.


\subsection{Spectrum and quantum geometry of the lattice model}
Let us now inspect more carefully the properties of the two-dimensional lattice model.
Introducing the Fourier transform of the sublattice-resolved fermionic annihilation $\left(A/B\right)_{\kappa,q}=N^{-1/2}\sum_je^{ij\kappa}\left(A/B_{j,q}\right)$, and the spinor $\Psi_{\kappa,q}=\left(A_{\kappa,q},B_{\kappa,q}\right)^T$, we find
\begin{equation}
    H_{\rm CI}=\sum_{\kappa,q}
    \Psi_{\kappa,q}^\dagger
    \left[
-t\left(1+\cos 2q\right)\sigma_x -t\sin 2q \sigma_y
+2t_1\sin\kappa\sigma_z+2t_2\cos\kappa\sigma_x 
    \right]
    \Psi_{\kappa,q},
    \label{eq:HCImomentumspace2d}
\end{equation}
and the resultant two bands have the spectrum
\begin{equation}
    E_{\kappa,q} = \pm 2\sqrt{\left(1+
    2\cos\kappa \,t_2/t\right)t^2\cos^2q
    +t_1^2\sin^2\kappa+ t_2^2\cos^2\kappa}.\label{eq:HCIspectrum2d}
\end{equation}
The bulk gap in the spectrum is $E_{\rm gap}=2\left| \left|M_+\right| - \left|M_-\right|\right|$.
When either $t_1=0$ or $t_2=0$, $\left|M_+\right|=\left|M_-\right|\equiv M$, the spectrum is gapless and has two anisotropic Dirac cones at 
$\left(\kappa,q\right)=\left(\frac{\pi}{2}/\frac{3\pi}{2},\frac{\pi}{2}\right)$
or 
$\left(\kappa,q\right)=\left(0/\pi,\frac{\pi}{2}\right)$,
respectively.
The Dirac cone velocity along the wire direction is $v=2t$ as before, whereas in the transverse direction it is $v_\perp=4M$.
We illustrate these points in Fig.~\ref{fig:touchingbandsdiraccones}.

\begin{figure}
\begin{centering}
\includegraphics[width=14cm]{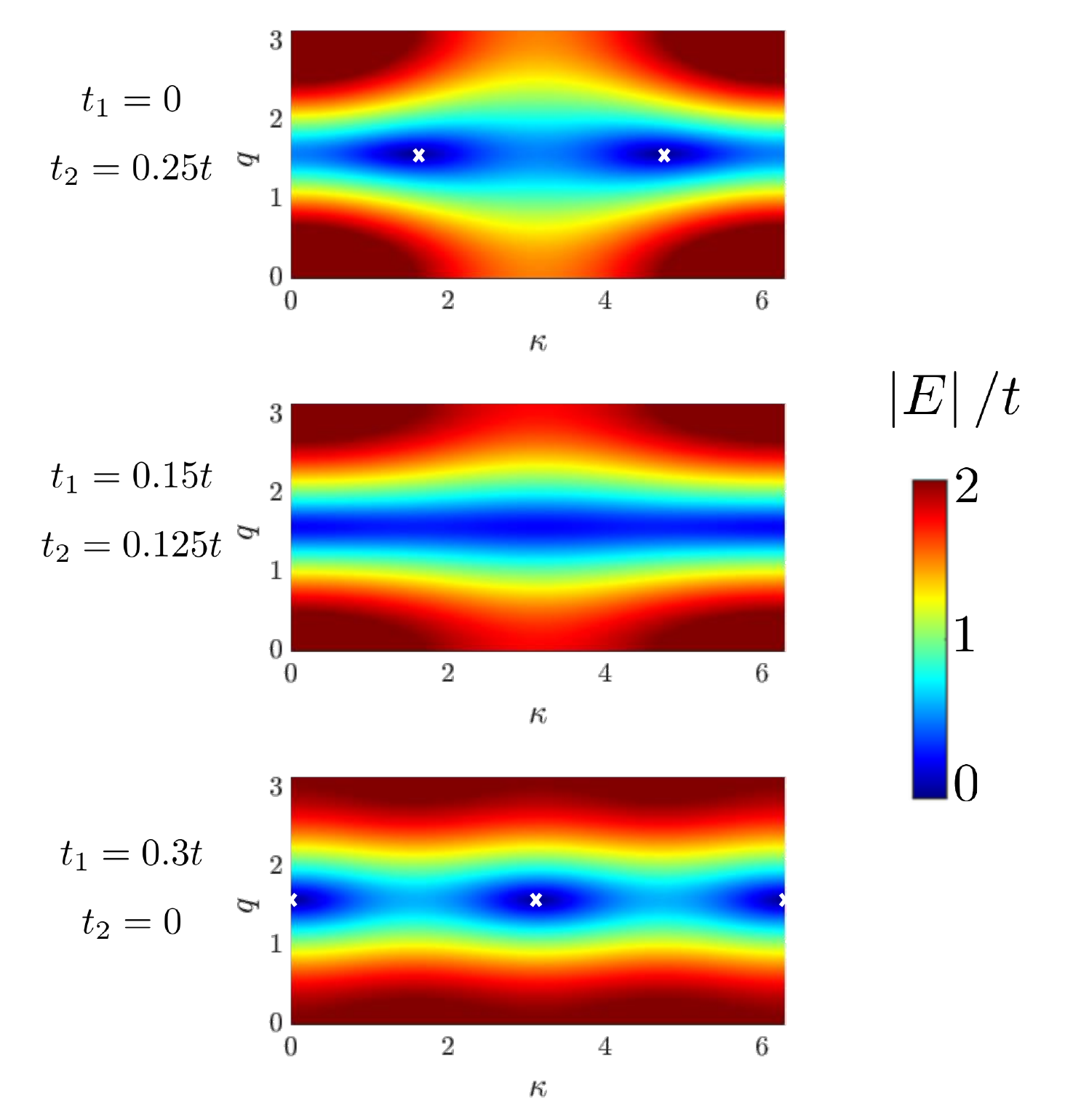}
\par\end{centering}
\caption{\label{fig:touchingbandsdiraccones} 
Illustration of the band structure at two gap closing points (top and bottom panel), and in between, where a gap remains open (middle panel).
The corresponding parameters are on the left of their respective panels.
As is shown, when either $t_1$ or $t_2$ vanish, the gap closes at two Dirac nodes, indicated by white x-marks.
The vanishing of either one of these interwire coupling corresponds to 
$\left|M_+\right|=\left|M_-\right|$.
Notice the dispersion around the Dirac nodes is highly dispersive in the $q$ direction, and rather flat in the $\kappa$ direction, as we discuss following Eq.~\eqref{eq:HCIspectrum2d}.
}
\end{figure}

We note that $H_{\rm CI}$ may be mapped onto an anisotropic version of the half-BHZ model~\cite{BHZmodelScience}.
This can be made clear by defining $k_x\equiv\kappa $, $k_y\equiv 2q$, and rotating the Pauli matrices around $\sigma_y$ by $\pi/2$.
This produces the momentum-resolved Hamiltonian
\begin{equation}
    h_{k_x,k_y}/\left(-t\right)=\frac{2t_1}{t}\sin k_x\sigma_x+\sin k_y \sigma_y+
    \left(1+\frac{2t_2}{t}\cos k_x+\cos k_y\right)\sigma_z. \label{eq:bhzdeformed}
\end{equation}
It is therefore not surprising that the topological properties of the two bands are similar to those of the half-BHZ model.

Let us now turn to explicitly examine the topological properties of the energy bands and their quantum geometry.
Without loss of generality we will focus on the valence band, spanned by the Bloch wavefunctions $\left| u_{\mathbf k}\right\rangle$, where as in our previous definition, $\mathbf{k}=\left(\kappa,2q\right)$.
Adopting similar conventions to Ref.~\cite{LedwithFCIgeometry}, we define the quantum geometrical tensor
\begin{equation}
    \eta_{\alpha\beta}\left({\mathbf k}\right)=\frac{\left\langle \partial_\alpha u_\mathbf{k}| \partial_\beta u_\mathbf{k} \right\rangle}{\left\langle u _{\mathbf{k}}|u _{\mathbf{k}}\right\rangle} 
    - 
    \frac{\left\langle \partial_\alpha u_\mathbf{k}\left|u_{\mathbf{k}}\right\rangle\left\langle u_{\mathbf{k}}\right| \partial_\beta u_\mathbf{k} \right\rangle}{\left\langle u _{\mathbf{k}}|u _{\mathbf{k}}\right\rangle^2},
    \label{eq:quantumgeometricaltensor}
\end{equation}
with the shorthand $\partial_\alpha = \frac{\partial}{\partial k_{\alpha}}$.
The Berry curvature $\Omega$ and Fubini-Study metric $g$ are then
\begin{equation}
    \Omega  = 2{\rm Im}\eta_{yx}, \,\,\,g_{\alpha\beta}={\rm Re}\eta_{\alpha\beta}.\label{eq:berrycurvatureandmetric}
\end{equation}

We will be interested in two so-called indicators of the band's susceptibility to hosting an FCI state.
The first is a measure of the Berry curvature fluctuations in the BZ,
\begin{equation}
    \sigma_\Omega = \sqrt{\int\frac{ d^2{\mathbf k}}{A}\left(A\frac{\Omega}{2\pi}-C\right)^2},\label{eq:berryfluctauationsrms}
\end{equation}
where $A$ is the area of the BZ and $C=\int d^2{\mathbf k}\frac{\Omega}{2\pi}$ is the Chern number of the band.
When the Berry curvature is a constant in the BZ $\sigma_\Omega=0$, it has been shown~\cite{parmesawaranberrycurvatureflatness} that the GMP algebra is exactly reproduced in the long-wavelength limit.
The second indicator is the so-called trace condition.
We define
\begin{equation}
    \Bar{T} = \int \frac{d^2{\mathbf k}}{A}\left({\rm tr} g - \left|\Omega\right|\right)\label{eq:tracecondition}
\end{equation}
as a measure of the saturation of the inequality ${\rm tr}g\geq\left|\Omega\right|$.
When the metric is $\mathbf{k}$-independent and this inequality is saturated the full GMP algebra can be recovered~\cite{RoytraceconditionFCI}.

When $\sigma_\Omega=0$ and $\Bar{T}=0$ (and the band is entirely flat) after projecting the interactions to the band of interest, one is thus essentially left with the physics of the lowest Landau level, which is of course ideally suited for a fractional Laughlin-like topological phase. 
There is also empirical evidence that even away from this ideal limit, it is beneficial to minimize these indicators to get a more FCI-friendly system.
In Ref.~\cite{FCIgeometrycorrelations} it was numerically demonstrated that the FCI many-body gap is correlated with small values $\sigma$ and $\Bar{T}$.
A similar trend was shown in Ref.~\cite{FCI_TBG_parker2021fieldtuned}, where a model of magic angle twisted bilayer graphene was studied.

\begin{figure}
\begin{centering}
\includegraphics[width=18cm]{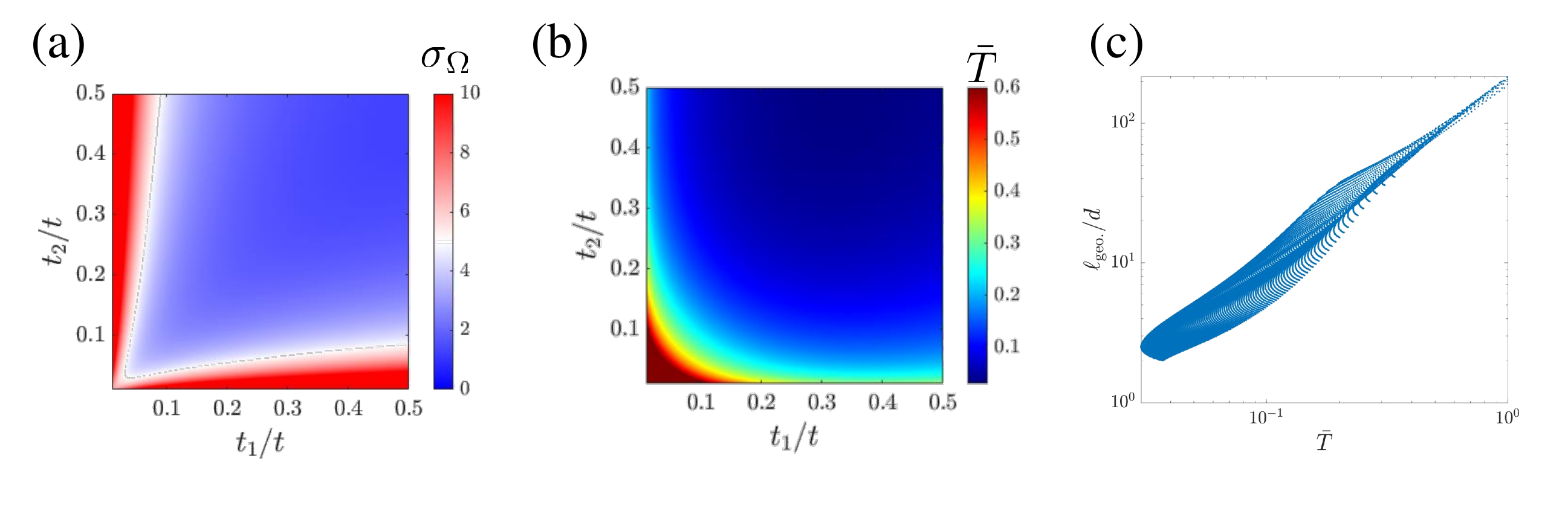}
\par\end{centering}
\caption{\label{fig:indicatorsgeometry} 
Quantum geometry indicators for FCI stabilization, calculated for the full two dimensional lattice model, Eq.~\eqref{eq:HCImomentumspace2d}.
(a)
Standard deviation of the Berry curvature $\sigma_\Omega$ [Eq.~\eqref{eq:berryfluctauationsrms}].
(b)
Violation of the trace condition [Eq.~\eqref{eq:tracecondition}].
As the two inter-wire coupling processes become of similar strength (and one of the mass terms dominates the other), the indicators generally become smaller, i.e., FCI should be more stable.
(c)
Correlation between the proxy length scale $\ell_{\rm geo.}$ [Eq.~\eqref{eq:geometriclength}], and trace condition violation $\Bar{T}$ (over the field of view of panel (b)). 
Each data point corresponds to a single ($t_1$ , $t_2$) coordinate in panel (b), where we calculate both $\Bar{T}$ and $\ell_{\rm geo.}$.
Clearly, the two are well correlated, justifying our focused attention on the former as an indicator for the latter.
}
\end{figure}

These so-called indicators are calculated for our proposed two-dimensional model as a function of the different inter-wire coupling strengths, see Fig.~\ref{fig:indicatorsgeometry}.
The clearest trend one observes is that these indicators are optimized when $\left|t_1\right|\to\left|t_2\right|$, or alternatively when $\left|M_+\right| \gg \left|M_-\right|$ (or vice versa).
As we will now demonstrate, this  also coincides with the minimization of the correlation length.
We note that as $t_1$ and $t_2$ become comparable with the interwire hopping $t$, the ideal quantum geometry as indicated by the trace condition (as seen Fig.~\ref{fig:indicatorsgeometry}) deviates slightly from the $\left|t_1\right|=\left|t_2\right|$ line.
In this regime, the full two dimensional band geometry plays an important role, and our coupled wires approach is also much less valid.

\subsection{Transverse correlation length}
Generically, $\left|t_1\right|\neq\left|t_2\right|$ and both $M_\pm$ are finite.
The physics along the transverse inter-wire direction can be mapped onto two copies of the Su-Schrieffer-Heeger (SSH) model~\cite{sshprl}: one copy is formed by right-movers on even wires coupled to left-movers on odd wires with alternating $M_+$ and $M_-$ hopping, and the second chain comprised by left-movers on even wires and the right-movers on odd wires.
This mapping provides us with two immediate results.
First, if $\left|M_+\right|=\left|M_-\right|$, the wires are critically coupled and the system remains gapless.
This was already understood from Eq.~\eqref{eq:HCIspectrum2d}.
This underscores that both $t_1$ and $t_2$ are required to manifest the Chern insulator phase.

\begin{figure}
    \includegraphics[width=10cm]{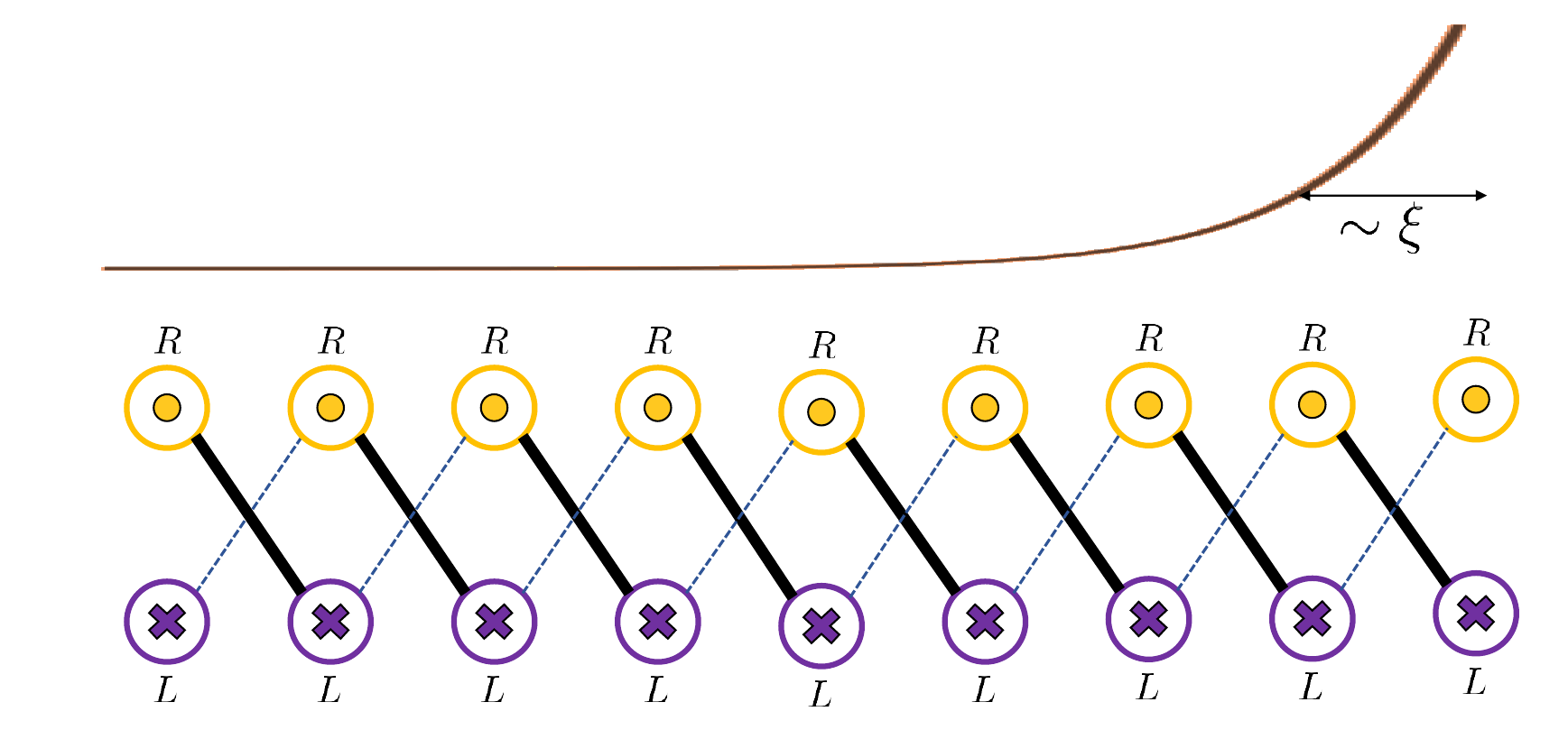}
    \centering
    \caption{\label{fig:sshrlmovers} 
    Coupling between the chiral modes of neighboring wires.
    Each wire hosts a right-moving mode ($R$, yellow) and a left-moving mode ($L$, purple).
    Adjacent wires are coupled by $M_+$ (thick black line) and $M_-$ (dashed blue line).
    The system is comprised of two decoupled zigzag chains with staggered couplings in each, analogous to the SSH model.
    The spatial extent of the localized edge-modes, $\xi$ [see Eq.~\eqref{eq:sshcorrelationlength}], is illustrated in the top part of the figure as an exponential decay in the transverse (interwire) direction.
}
\end{figure}

On the other hand, if the mass terms have different magnitudes, one still finds counter-propagating edge modes in the system (corresponding to the SSH edge states), except in this more general case they are \textit{not entirely localized} on the outermost wires.
Instead, their support decays in the transverse direction with a correlation length (in units of the inter-wire separation)~\cite{sshcorrelationlength}
\begin{equation}
    \xi=\frac{2}{\log\frac{{\rm max}\left\{\left|M_+\right|,\left|M_-\right|\right\}}{{\rm min}\left\{\left|M_+\right|,\left|M_-\right|\right\}}}
    =
    \frac{2}{\log \frac{\left|t_1\right|+\left|t_2\right|}
    {\left|\left|t_1\right|-\left|t_2\right|\right|}}
    .
    \label{eq:sshcorrelationlength}
\end{equation}

This finite correlation length provides a clear important distinction from previous works regarding wire constructions of quantum Hall states~\cite{wireconstructionKanePrl,wireconstructionkaneteoPrb}.
In these constructions, the chiral edge modes were always localized on one wire, i.e., the effective correlation length was $\xi=0$.
We note that at a certain regime of parameters in one of the models presented in Ref.~\cite{SagiFCIconstruction}, a finite extent of the chiral edge modes is made possible, yet this possibility and its potential importance remained unexplored.
In the the type of model we consider, tuning the coupling parameters $t_1$ and $t_2$ provides control over this localization.
This becomes crucial to our analysis and understanding of the fractional phases.

\section{Relating quantum geometry and the correlation length}

Consider the continuum limit of the coupled wires model, expanded in small momentum along the wire direction (small $q$),

\begin{equation}
    H_{\rm cont.}=\sum_{k,q}\left[vq\left(\psi_{R,k,q}^{\dagger}\psi_{R,k,q}-\psi_{L,k,q}^{\dagger}\psi_{L,k,q}\right)+\left(M_{+}e^{ikd}\psi_{R,k,q}^{\dagger}\psi_{L,k,q}+M_{-}e^{-ikd}\psi_{R,k,q}^{\dagger}\psi_{L,k,q}+{\rm h.c.}\right)\right], \label{eq:Hcontforauntumlength}
\end{equation}
where $\psi_{R/L,k,q}$ is a right/left-moving fermionic annihilation operator at momentum $k,q$ ($k$ is the momentum in the transverse direction), $M_{\pm}$ are the different interwire couplings, and $d$ is the interwire distance (which is re-introduced for clarity).
By defining the spinor $\Psi_{k,q}=\left(\psi_{R,k,q},\psi_{L,k,q}\right)^{T}$, one may rewrite this Hamiltonian,
\begin{equation}
    H=\sum_{k,q}\Psi_{k,q}^{\dagger}h_{k,q}\Psi_{k,q},
\end{equation}
with
\begin{align}
h_{k,q} & =vq\sigma_{z}+\left(M_{+}+M_{-}\right)\cos kd\sigma_{x}+\left(M_{+}-M_{-}\right)\sin kd\sigma_{y}\nonumber\\
 & \equiv vq\sigma_{z}+a\cos kd\sigma_{x}+b\sin kd\sigma_{y}.\label{eq:hkq_forgeometriclength}
\end{align}
The spectrum of the Hamiltonian is $\epsilon_{k,q}=\pm\sqrt{\left(vq\right)^{2}+a^{2}\cos^{2}kd+b^{2}\sin^{2}kd}\equiv\pm E_{k,q}.$
We denote the eigen-wavefunction of the bottom band as $|u_{k,q}\rangle$, with their explicit form
\begin{equation}
    |u_{k,q}\rangle=\begin{pmatrix}\sqrt{\frac{E_{k,q}+vq}{2E_{k,q}}}\\
\frac{a\cos kd-ib\sin kd}{\sqrt{2E_{k,q}\left(E_{k,q}+vq\right)}}
\end{pmatrix}.\label{eq:ukq_wavefunction}
\end{equation}

To understand how tuning the model affects the quantum geometry it is useful to inspect the following quantity,
\begin{equation}
    \ell_{\rm geo.}=4 \int\frac{dk}{2\pi}{\rm tr} g \left(k,q=0\right),\label{eq:geometriclength}
\end{equation}
which is a length scale, associated with the spread of the maximally localized Wannier functions along the transverse direction.
Clearly, it is also one of the components that make up the trace condition, and it is in fact the component most influenced by $M_{\pm}$ tuning.
We have illustrated in Fig.~\ref{fig:indicatorsgeometry}c the direct correlation between this length scale and the trace condition violation $\Bar{T}$.
Our definition of $\ell_{\rm geo.}$ is in analogy to Ref.~\cite{Vanderbiltmaximmalylocalized}, that showed that the trace-condition violation quantifies the spatial extent of the maximally localized Wannier functions. 

From straightforward calculation using Eqs.~\eqref{eq:quantumgeometricaltensor} and~\eqref{eq:ukq_wavefunction}, we find
\begin{equation}
    \eta_{kk}\left(k,q=0\right)=\frac{d^{2}}{4}\frac{1-\delta\cos2kd}{1+\delta\cos2kd},
\end{equation}
\begin{equation}
    \eta_{qq}\left(k,q=0\right)=\frac{d^2}{4}
    \left(\frac{v}{d M_+}\right)^2
    \frac{1}{1+e^{-\frac{4d}{\xi}}}\frac{1}{1+\delta\cos2kd},
\end{equation}
with $\delta=\frac{2M_{+}M_{-}}{M_{+}^{2}+M_{-}^{2}}$, and without loss of generality we assumed $M_+>M_-$. 
Notice that in the maximally-chiral limit, when one of the interwire terms overwhelms the other, $\delta\to0$, and the above quantities are ``flat'', i.e., independent of $k$.
We finally recover $\ell_{\rm geo.}$, and relate it to the transverse correlation length (see Fig.~\ref{fig:ellgeometric}),
\begin{equation}
    \ell_{\rm geo.}=d\frac{3+\left[1+\left(\frac{v}{dM_{+}}\right)^{2}\right]e^{\frac{4d}{\xi}}}{e^{\frac{4d}{\xi}}-1}.\label{eq:lgeometricresult}
\end{equation}

Let us examine two interesting simple limits.
When the correlation length is vanishingly small as compared to the inter-wire separation, i.e., one of the mass terms dominates the other,
\begin{equation}
     \ell_{\rm geo.} \left( \xi \ll d\right) \approx d\left[1+\left(\frac{v}{dM_{+}}\right)^{2}\right]. \label{eq:lgeometricSMALL}
\end{equation}
This is the lower bound for $\ell_{\rm geo.}$, indicating that due to the topological non-triviality of the band, the Wannier functions of this band cannot be localized on a single wire (obstruction to the so-called atomic limit).

In the opposite limit, where the correlation length greatly exceeds the interwire separation, $\xi \gg d$,
\begin{equation}
     \ell_{\rm geo.} \left( \xi \gg d\right) \approx  
     \xi\left[1+\left(\frac{v}{2dM_{+}}\right)^{2}\right]+d\left[2\left(\frac{v}{2dM_{+}}\right)^{2}-1\right]
     \approx \xi\left[1+\left(\frac{v}{2dM_{+}}\right)^{2}\right]. \label{eq:lgeometricBIG}
\end{equation}
We thus find that our defined ``spread function'' $\ell_{\rm geo.}$ starts out being of order $\sim d$ when $\xi$ is very small (near optimal chiralness), and as the correlation length grows, $\ell_{\rm geo.}\propto\xi$. 
Relating this spread function to the correlation length $\xi$ analytically establishes the connection of the latter to quantum geometry and to the extent to which the trace condition is violated.

We also briefly mention that calculation of the Berry curvature at
optimal chiralness, $\xi\to0$ (without loss of generality, $M_{-}=0$),
\begin{equation}
    \Omega_{xy}\left(\xi\to0\right)=\frac{d}{2}\frac{v/M_{+}}{\left[\left(\frac{vq}{M_{+}}\right)^{2}+1\right]^{3/2}},
\end{equation}
is also completely flat in the $k$-direction, once more indicating quantum geometry which is more favorable towards FCI stabilization, i.e., the low variance of the Berry curvature in the BZ.
\begin{figure}
    \includegraphics[width=10cm]{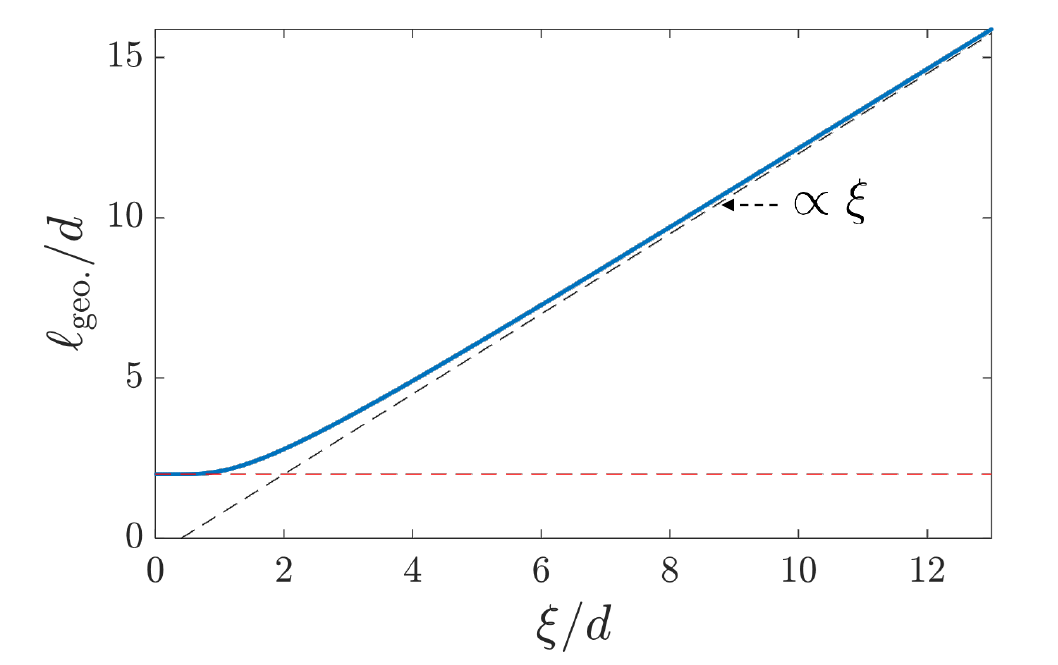}
    \centering
    \caption{\label{fig:ellgeometric} 
    Dependence of the trace condition component $\ell_{\rm geo.}$ [Eq.~\eqref{eq:geometriclength}] on the correlation length $\xi$, as calculated from Eq.~\eqref{eq:lgeometricresult} (blue).
    Red and black dashed lines show the small [Eq.~\eqref{eq:lgeometricSMALL}] and large [Eq.~\eqref{eq:lgeometricBIG}] $\xi/d$ asymptotics, respectively.
    In this plot, we set $v/\left(dM_+\right)=1$
}
\end{figure}

\section{Multi-particle interactions}
In the main text, we were mainly concerned with studying certain many-body scattering operators, their scaling dimension, and the phase diagram of our coupled wires construction in their presence.
These are
\begin{equation}
    {\cal O}^j_{\rm FCI} \sim g_{\rm FCI} 
    \left({\cal O}_{j,{\rm bs}}\right)^p
    \left({\cal O}_{j+1,{\rm bs}}\right)^p
    \psi_{j,R}^\dagger \psi_{j+1,L}
    +{\rm h.c.} \label{eq:OFCImainsupp}
\end{equation}
\begin{equation}
    {\cal O}^j_{\rm aFCI} \sim g_{\rm aFCI} 
    \left({\cal O}_{j,{\rm bs}}\right)^p
    \left({\cal O}_{j+1,{\rm bs}}\right)^p
    \psi_{j+1,R}^\dagger \psi_{j,L}
    +{\rm h.c.} \label{eq:OaFCImainsupp}
\end{equation}
\begin{equation}
    {\cal O}^j_{\rm CDW} = g_{\rm CDW}\left({\cal O}_{j,{\rm bs}}\right)^{2p+1}
    +{\rm h.c.}\label{eq:OCDWmainsupp}
\end{equation}
\begin{figure}
    \includegraphics[width=18cm]{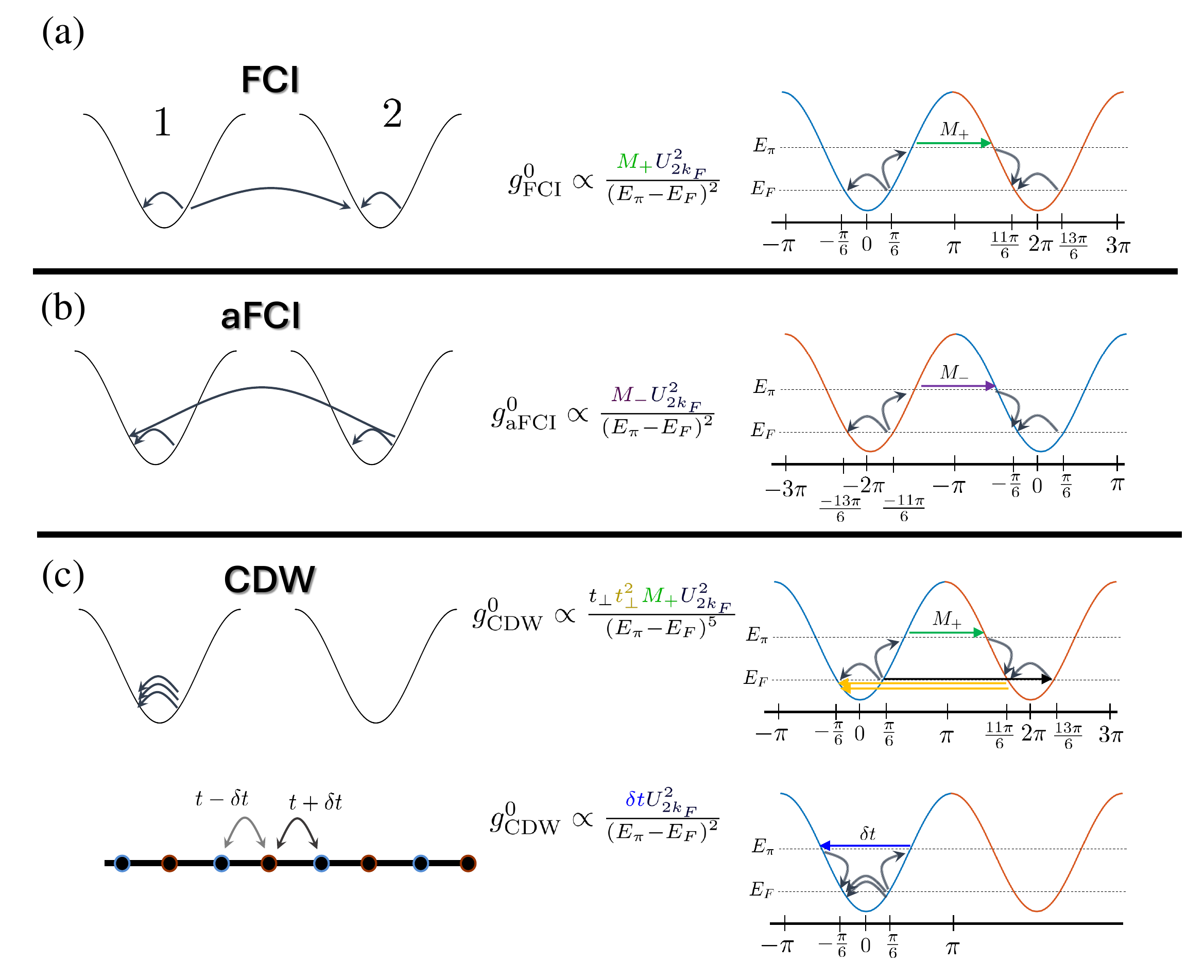}
    \centering
    \caption{\label{fig:perturbationcouplingconstants} 
    Origin of many-body scattering processes in the interaction Hamiltonian at filling $\nu=1/3$.
    (a)
    A process that corresponds to the FCI stabilization.
    The left bands, 1 and 2, correspond to the dispersion in two neighboring wires.
    On the right, we shift the dispersion of wire 2 (red) by $2\pi$ with respect to wire 2 (blue).
    This is allowed by the periodicity of the Brillouin zone.
    The energies $E_\pi$ and $E_F$ are the chemical potentials corresponding to half-filling and the Fermi energy, respectively.
    The arrows on the right correspond to interactions at momentum $2k_F$ with strength $U_{2k_F}$ (gray-blue lines), and the $M_+$ process (green).
    Notice $M_+$ conserves momentum modulo $\pi$ and not $2\pi$, since it originates in interwire coupling, which is modulated between adjacent sites.
    (b)
    The process corresponding to aFCI.
    On the right, we perform a similar shift to (a), except here the shift is by $-2\pi$ instead of $2\pi$, hence the colors are reversed.
    The relevant momentum-$\pi$ process here is $M_-$ (purple).
    (c)
    The interaction stabilizing the CDW phase we discuss in this work.
    The top right panel features 3 additional $t_\perp$ hoppings (2 in yellow, one in black), which couple carriers of the same chirality between neighboring wires [see Eq.~\eqref{eq:supptperp}].
    The bottom right panel demonstrates a lower order process, which is enabled by a slight modulation of the intrawire hopping strength by $\delta t$.
    Due to the modulation, this hopping is allowed to carry momentum of $\pi$.
}
\end{figure}
For the sake of completeness, let us discuss here how these high-order scattering processes may be generated through perturbation theory, as one considers interactions, interwire coupling, and overall momentum conservation.
We focus here on the $\nu=1/3$ case for concreteness.

The process related to FCI stabilization, described by ${\cal O}^j_{\rm FCI}$, is shown in Fig.~\ref{fig:perturbationcouplingconstants}a.
By an arbitrary momentum shift of one wires dispersion relative to its neighbor by $2\pi$, on readily sees the processes involved.
Namely, two $2k_F$ scattering processes of strength $U_{2k_F}$, combined with one interwire hopping $\propto M_+$.
Although it seems $M_+$ leads to violation of momentum conservation, due to the modulation of the interwire hoppings between adjacent intrawire sites, this process carries a momentum of $\pm\pi$, enabling the conservation of momentum.

The analogous process for the aFCI, ${\cal O}^j_{\rm aFCI}$, appears in panel b of Fig.~\ref{fig:perturbationcouplingconstants}.
Its analogy to the FCI is made obvious by shifting relative momentum between the wires by $-2\pi$ instead of $+2\pi$, illustrating how now $M_-$ is required to facilitate the scattering.
The upshot here is that  the initial coupling constants of these processes, $g_{\rm FCI}^0$ and $g_{\rm aFCI}^0$ are identical up to the transmutation $M_+ \leftrightarrow M_-$.
This difference lies at the heart of relating the competition between the two to the quantum geometric properties of the hosting bands.

Lastly, we comment on the CDW process, ${\cal O}^j_{\rm CDW}$, illustrated in Fig,~\ref{fig:perturbationcouplingconstants}c.
Following the same procedure, the initial coupling $g^0_{\rm CDW}$ appears to potentially be of much higher order, as it contains an additional $t_\perp^3$ factor relative to the previous two processes.
This traces back to the fact that one requires momentum-$\pi$ carrying processes to establish momentum conservation.
As the lattice model stands, only interwire hopping is able to achieve this, hence the necessity of the second wire participating in this game.
However, by slightly modifying the model, as to modulate the intrawire hopping $t$ between adjacent sites, i.e., $t\to t\pm\delta t$, one immediately enables the lower order process shown in the bottom of Fig.~\ref{fig:perturbationcouplingconstants}.

\section{Weak coupling RG}

To study the competition between the three different multi-particle backscattering terms, which correspond to different correlated phases potentially stabilized in the system, we employ the perturbative renormalization group (RG) approach.
At the level of weak-coupling analysis, the competition is sufficiently well captured by a model comprised of just two neighboring wires.
Within this framework, the effect of electron-electron interactions on the competition between the phases can be well understood.

Let us consider the Hamiltonian density
\begin{equation}
    {\cal H}= {\cal H}_0 + {\cal H}_{\rm CDW} + {\cal H}_{\rm FCI},
\end{equation}
with 
\begin{equation}
    {\cal H}_0 = \frac{1}{2\pi}\sum_{i=\rho,\sigma}u_{i}
    \left[K_i^{-1} \left(\partial_x \phi_i\right)^2
    +K_i \left(\partial_x \theta_i\right)^2
    \right],
\end{equation}
\begin{equation}
    {\cal H}_{\rm CDW} = \frac{g_{\rm CDW}}{2\pi^2}
    \cos\left(m\sqrt{2}\phi_{\rho}\right)
    \cos\left(m\sqrt{2}\phi_{\sigma}\right)
    +\frac{g_{\phi}}{2\pi^{2}}\cos\left(\sqrt{8}\phi_{\sigma}\right),
\end{equation}
\begin{equation}
    {\cal H}_{\rm FCI} = \frac{g_{{\rm FCI}}}{2\pi^{2}}\cos\left(\sqrt{2}\theta_{\sigma}+m\sqrt{2}\phi_{\rho}\right)
    +\frac{g_{{\rm aFCI}}}{2\pi^{2}}\cos\left(\sqrt{2}\theta_{\sigma}-m\sqrt{2}\phi_{\rho}\right)
    +\frac{\tilde{V}}{2\pi}
    \left(\partial_x\phi_\rho\partial_x\theta_\sigma+\partial_x\theta_\rho\partial_x\phi_\sigma\right).
\end{equation}
Here the two bosonic sectors $\rho,\sigma$ correspond to different combinations of the fields on the two wires labeled $1,2$, e.g.,
$\phi_{\rho/\sigma}=1/\sqrt{2}\left(\phi_1\pm\phi_2\right)$.

Within the unperturbed ${\cal H}_0$, $g_{\rm CDW}$ has the scaling dimension $d_{\rm CDW}$, and the two FCI terms have the same scaling dimension $d_{\rm FCI}$, where
\begin{equation}
    d_{\rm CDW} = \frac{m^2}{2}K_\rho+\frac{m^2}{2}K_\sigma,
\end{equation}
\begin{equation}
    d_{\rm FCI} = \frac{m^2}{2}K_\rho+\frac{1}{2}K_\sigma^{-1}.
\end{equation}
Clearly, $K_\rho$, which corresponds to the total charge sector and thus is expected to be rather small in the case of strong repulsive interactions, will not play any meaningful role in the competition of these different backscattering terms, as all three depend on it in the exact same way.
Instead however, $K_{\rho}$ controls the transition between a gapless metallic phase for $K_{\rho}\lesssim 1$ (weak repulsive interactions), and the strong coupling phase of one or several of the different $g_i$, mandating $K_{\rho}\ll 1$, i.e., strong repulsive interactions.

Conversely, one immediately notices that $K_{\sigma}$ directly controls the competition between the CDW phase requiring sufficiently small $K_\sigma$, and the FCI terms which favor large values of $K_\sigma$.
The magnitude of $K_\sigma$ can be estimated by considering an interwire density-density interaction term $\frac{V_{12}}{\pi} \partial_x\phi_1 \partial_x \phi_2$, intra-wire Luttinger parameter $K$, and  the effective intra-wire Fermi velocity $v$.
As a function of these parameters we may express $K_{\rho/\sigma}=K/\sqrt{1\pm\frac{V_{12}K}{v}}$.
Assuming the intrawire $K$ is determined by a single density-density interaction $V_0$, and the bare Fermi velocity $v_F$, we may estimate
\begin{equation}
    K_\sigma = K\sqrt{\frac{v_F+V_0}{v_F+V_0-V_{12}}}.
\end{equation}
Thus, $K_\sigma$ is expected to be large if the interwire repulsion is comparable to, or even stronger than the intrawire repulsion $V_0$.

The term in ${\cal H}_{\rm CDW}$ proportional to $g_\phi$ originates in large-momentum transfer interactions between the two adjacent wires, i.e., $\psi^\dagger_{1,R}\psi_{1,L}\psi^\dagger_{2,L}\psi_{2,R}$.
It stabilizes a system-wide CDW by favoring the alignment of the local intra-wire CDWs to each other, such that a minima in the density in one wire tends to align to a maxima in its interacting neighbors.

Finally, let us address the seemingly peculiar $\tilde{V}$ interaction.
Since it is odd in $\theta_i$ fields, it explicitly breaks the time-reversal symmetry.
Its microscopic origin may come from the same time-reversal symmetry breaking which facilitated the formation of a Chern insulator (and thus differentiated also between $g_{\rm FCI}$ and $g_{\rm aFCI}$).
Alternatively, as we will show below, it is also directly generated at low energies when $g_{\rm FCI} \neq g_{\rm aFCI}$.

We derive the RG equations using the standard operator product expansion (OPE)~\cite{cardy_1996}. 
We parametrize the flowing short-distance cutoff as $\alpha=\alpha_0 e^\ell$, where in each RG step $\ell$ increases incrementally.
For the sake of simplicity, we neglect the differences between the velocities in different sectors, which impact the RG flow only in higher-orders than the ones considered.
We henceforth set $u_i\approx u$.

Defining dimensionless coupling constants $y_i\equiv g_i / \left(\pi u\right)$,
we find the following set of RG equations,
\begin{equation}
\begin{aligned}\label{eq:RGALLsupp}
    \frac{d}{d\ell}y_{{\rm FCI}}&=\left(2-d_{{\rm FCI}}+\frac{m}{2}K_{\rho}K_{\sigma}^{-1}\tilde{V}\right)y_{{\rm FCI}},\\
    \frac{d}{d\ell}y_{{\rm aFCI}}&=\left(2-d_{{\rm FCI}}-\frac{m}{2}K_{\rho}K_{\sigma}^{-1}\tilde{V}\right)y_{{\rm aFCI}},\\
    \frac{d}{d\ell}y_{{\rm CDW}}&=\left(2-d_{{\rm CDW}}\right)y_{{\rm CDW}},\\
    \frac{d}{d\ell}y_{\phi}&=\left(2-2K_{\sigma}\right)y_{\phi},\\
    \frac{d}{d\ell}K_{\rho}^{-1}&=\frac{m^{2}}{2}\left(y_{{\rm FCI}}^{2}+y_{{\rm aFCI}}^{2}+y_{{\rm CDW}}^{2}\right),\\
    \frac{d}{d\ell}K_{\sigma}&=\frac{1}{2}\left(y_{{\rm FCI}}^{2}+y_{{\rm aFCI}}^{2}\right)-K_{\sigma}^{2}\left(\frac{m^{2}}{2}y_{{\rm CDW}}^{2}+2y_{\phi}^{2}\right),\\
    \frac{d}{d\ell}\tilde{V}&=m\left(y_{{\rm FCI}}^{2}-y_{{\rm aFCI}}^{2}\right).
\end{aligned}
\end{equation}

The relationship between the ``proper'' FCI and the anti-FCI terms is now somewhat clarified by the RG equations.
At the level of weak-coupling, the competition is captured by the $\tilde{V}$ interaction discussed above.
This interaction (with a positive sign) directly aids the flow of $y_{\rm FCI}$ to strong coupling.
However the growth of $\tilde{V}$ itself is \textit{severely impeded by the mere presence of the counter term} $y_{\rm aFCI}$.
Thus, the presence of the latter imposes a burden on the possibility of stabilizing the FCI phase.
With similar reasoning, one observes that the CDW and the two FCI terms act in opposing ways on the flow of $K_\sigma$, which was shown above to be the most pertinent one for this specific competition.

\subsection{Derivation example}
Let us demonstrate our derivation of the RG equations, by considering the most non-trivial part, i.e., the contribution of $\Tilde{V}$ to the beta function of, e.g., $y_{\rm FCI}$.
Generally, the second order beta functions in 1+1d are written as
\begin{equation}
    \frac{d}{d\ell} y_k = \left(2-d_k\right)y_k - c_{ijk} y_i y_j, \label{eq:RGbetafunction}
\end{equation}
where $d_k$ is the scaling dimension of the operator corresponding to $y_k$, and summation over repeated indices is implied.
The coefficient $c_{ijk}$ can be identified from the OPE of the operators $O_{i/j}$ with the corresponding coupling constants $y_{i/j}$,
\begin{equation}
    :O_i\left({\bf z}_1\right)::O_j\left({\bf z}_2\right):
    =
    \frac{c_{ijk}}{\left|{\bf z}_1-{\bf z}_2\right|^{d_i + d_j - d_k}}
    :O_k\left(\frac{{\bf z}_1+{\bf z}_2}{2}\right):.
    \label{eq:normalorderedope}
\end{equation}
Therefore, we examine the following OPE,
\begin{align}
    I_{y_{\rm FCI},\Tilde{V}}&=
    :\nabla\phi_{\rho}\nabla\theta_{\sigma}::\cos\left(\sqrt{2}\theta_{\sigma}+m\sqrt{2}\phi_{\rho}\right):
    \nonumber\\
    &=
    \frac{1}{2}:\nabla\phi_{\rho}\nabla\theta_{\sigma}:\sum_{n=0}^{\infty}\frac{i^{n}}{n!}\sum_{k=0}^{n}\begin{pmatrix}n\\
k
\end{pmatrix}:\left(\sqrt{2}\theta_{\sigma}\right)^{k}\left(m\sqrt{2}\phi_{\rho}\right)^{n-k}:+{\rm h.c.}
\end{align}
We now need to start contracting the $\theta_\sigma$ fields and the $\phi_\rho$ fields.
One needs to ``choose'' out of $k$ terms for the former, and out of $n-k$ terms for the latter.
Thus,
\begin{align}
    I_{y_{\rm FCI},\Tilde{V}}&=2m\times\frac{1}{2}\sum_{n=0}^{\infty}\frac{i^{n}}{n!}\sum_{k=0}^{n}\begin{pmatrix}n\\
k
\end{pmatrix}k\left(n-k\right)\left\langle \nabla\theta_{\sigma}\theta_{\sigma}\right\rangle \left\langle \nabla\phi_{\rho}\phi_{\rho}\right\rangle :\left(\sqrt{2}\theta_{\sigma}\right)^{k-1}\left(m\sqrt{2}\phi_{\rho}\right)^{n-k-1}:+{\rm h.c.}\nonumber\\
&=
2m\times\frac{1}{2}\sum_{n=0}^{\infty}\frac{i^{n}}{n!}\sum_{k=0}^{n}\begin{pmatrix}n\\
k
\end{pmatrix}k\left(n-k\right)\left[-\frac{K_{\sigma}^{-1}}{2}\frac{{\bf z_1-z_2}}{\left|{\bf z_1-z_2}\right|^{2}}\right]\left[-\frac{K_{\rho}}{2}\frac{{\bf z_1-z_2}}{\left|{\bf z_1-z_2}\right|^{2}}\right]:\left(\sqrt{2}\theta_{\sigma}\right)^{k-1}\left(m\sqrt{2}\phi_{\rho}\right)^{n-k-1}:+{\rm h.c.}
\nonumber\\
&=
\left[\frac{m}{2}\frac{K_{\rho}K_{\sigma}^{-1}}{\left|{\bf z_1-z_2}\right|^{2}}\right]\times
\frac{1}{2}\sum_{n=0}^{\infty}\frac{i^{n}}{n!}\sum_{k=0}^{n}\begin{pmatrix}n\\
k
\end{pmatrix}k\left(n-k\right):\left(\sqrt{2}\theta_{\sigma}\right)^{k-1}\left(m\sqrt{2}\phi_{\rho}\right)^{n-k-1}:+{\rm h.c.}
\nonumber\\
&=
\left[\frac{m}{2}\frac{K_{\rho}K_{\sigma}^{-1}}{\left|{\bf z_1-z_2}\right|^{2}}\right]\times
\frac{1}{2}\sum_{n=0}^{\infty}\frac{i^{n}}{n!}\sum_{k=0}^{n}\frac{n\left(n-1\right)\left(n-2\right)!}{\left(k-1\right)!\left(n-2-\left(k-1\right)\right)!}:\left(\sqrt{2}\theta_{\sigma}\right)^{k-1}\left(m\sqrt{2}\phi_{\rho}\right)^{n-2-\left(k-1\right)}:+{\rm h.c.}
\nonumber\\
&=
\left[\frac{m}{2}\frac{K_{\rho}K_{\sigma}^{-1}}{\left|{\bf z_1-z_2}\right|^{2}}\right]\times
\frac{1}{2}\sum_{n=0}^{\infty}\frac{i^{n}}{n!}n\left(n-1\right)\sum_{k=0}^{n}\begin{pmatrix}n-2\\
k-1
\end{pmatrix}:\left(\sqrt{2}\theta_{\sigma}\right)^{k-1}\left(m\sqrt{2}\phi_{\rho}\right)^{n-2-\left(k-1\right)}:+{\rm h.c.}
\nonumber\\
&=
\left[\frac{m}{2}\frac{K_{\rho}K_{\sigma}^{-1}}{\left|{\bf z_1-z_2}\right|^{2}}\right]\times
\frac{1}{2}\sum_{n=0}^{\infty}\frac{i^{n-2}i^{2}}{\left(n-2\right)!}:\left(\sqrt{2}\theta_{\sigma}+m\sqrt{2}\phi_{\rho}\right)^{n-2}:+{\rm h.c.}
\nonumber\\
&=
-\left[\frac{m}{2}\frac{K_{\rho}K_{\sigma}^{-1}}{\left|{\bf z_1-z_2}\right|^{2}}\right]\times
:\cos\left(\sqrt{2}\theta_{\sigma}+m\sqrt{2}\phi_{\rho}\right):.
\end{align}
We thus identify 
\begin{equation}
    c_{\tilde{V},y_{\rm FCI},y_{\rm FCI}} = \frac{m}{2}K_{\rho}K_{\sigma}^{-1}.
\end{equation}

\subsection{Alternative definitions}\label{appsec:yfz}
As in the main text, it is convenient to re-define
\begin{equation}
    y_{\rm F}^2 = y_{\rm FCI}^2 + y_{\rm aFCI}^2,
\end{equation}
\begin{equation}
    y_{\rm F}^2 z = y_{\rm FCI}^2 - y_{\rm aFCI}^2.
\end{equation}
Notice $z \in \left[0 , 1\right]$, where $z=1$ corresponds to the maximally chiral $\xi=0$ case.
With these alternative representations, one finds
\begin{equation}
\begin{aligned}
    \frac{d}{d\ell}y_{{\rm F}}&=\left(2-d_{{\rm FCI}}+\frac{m}{2}K_{\rho}K_{\sigma}^{-1}\tilde{V}z\right)y_{{\rm F}},\\
    \frac{d}{d\ell}z&=mK_{\rho}K_{\sigma}^{-1}\tilde{V}\left(1-z^{2}\right),\\
    \frac{d}{d\ell}y_{{\rm CDW}}&=\left(2-d_{{\rm CDW}}\right)y_{{\rm CDW}},\\
    \frac{d}{d\ell}y_{\phi}&=\left(2-2K_{\sigma}\right)y_{\phi},\\
    \frac{d}{d\ell}K_{\rho}^{-1}&=\frac{m^{2}}{2}\left(y_{{\rm F}}^{2}+y_{{\rm CDW}}^{2}\right),\\
    \frac{d}{d\ell}K_{\sigma}&=\frac{1}{2}y_{{\rm F}}^{2}-K_{\sigma}^{2}\left(\frac{m^{2}}{2}y_{{\rm CDW}}^{2}+2y_{\phi}^{2}\right),\\
    \frac{d}{d\ell}\tilde{V}&=m z y_{\rm F}^2.
\end{aligned}
\end{equation}
From this form of the RG equations, it becomes clear that $z>0$ aids the growth of $y_{\rm F}$ to strong coupling, both directly and by generating (or enhancing) the time-reversal odd interaction $\tilde{V}$.

\subsection{Additional phase diagrams for different $z$}
We illustrate the full evolution of the phase diagram, as obtained in Figure 2 of the main text, as a function of ``deteriorating'' quantum geometry.
This is shown in Fig.~\ref{fig:supp_extended_phase}.
As anticipated, the region where the FCIs are stabilized shrinks, as the so-called aFCI seed becomes larger, i.e., $z$ becomes smaller, and $\ell_{\rm geo.}$ moves further away from ts optimal value.

\begin{figure}
    \includegraphics[width=18cm]{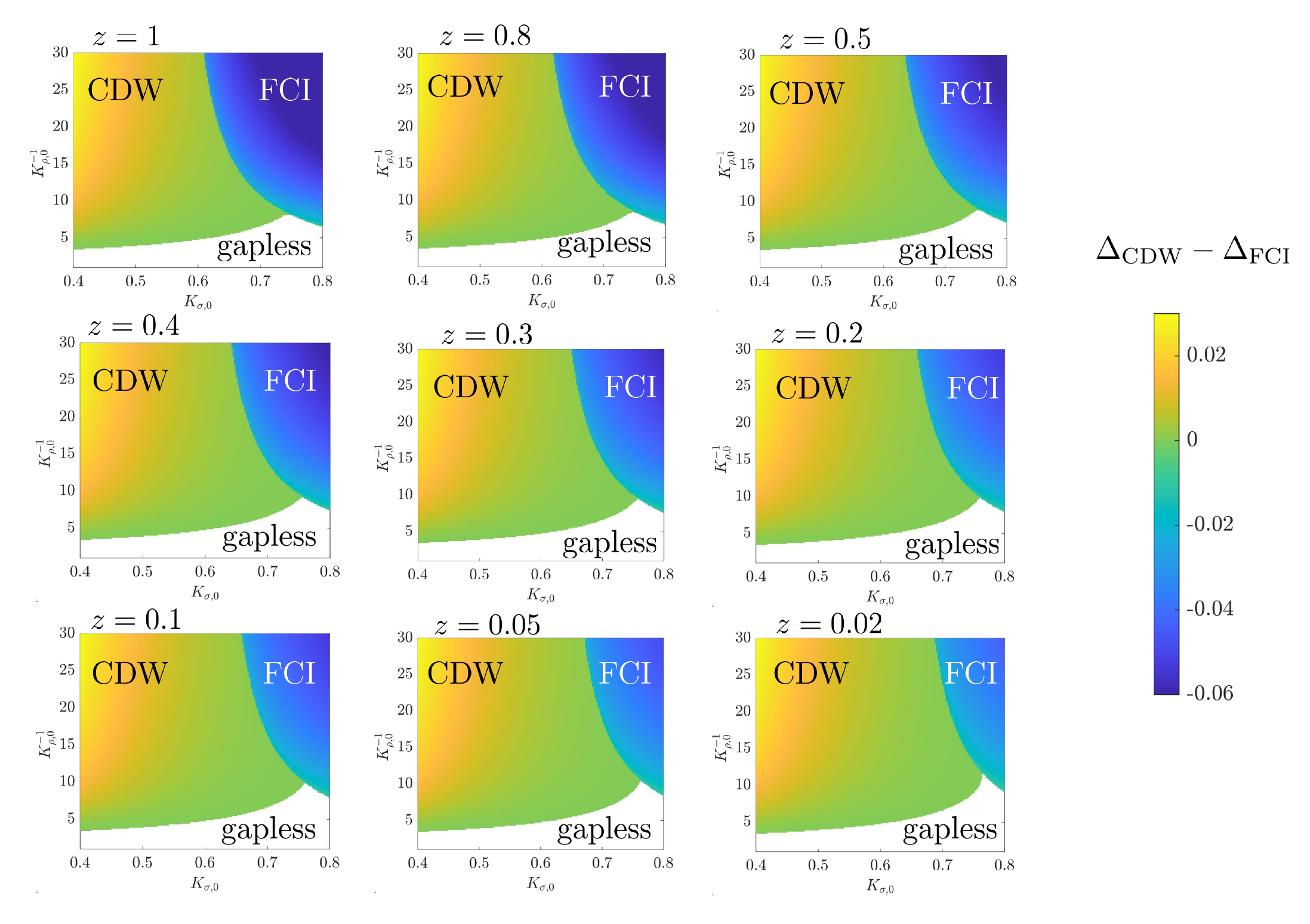}
    \centering
    \caption{\label{fig:supp_extended_phase} 
    Phase diagrams obtained by the RG flow, Eq.~\eqref{eq:RGALLsupp}, for different values of $z=\tanh\frac{2d}{\xi_{\rm topo.}}$, indicated above each panel.
    We plot the difference between the gap proxies $\Delta_{\rm CDW}-\Delta_{\rm FCI}$ for deteriorating quantum geometry.
    Here $m=3$ and initial conditions are $y_{F,0}=0.03$, $y_{\rm CDW,0}=0.08$, $y_{\phi,0}=0.1$, $\tilde{V}_0=0$.
}
\end{figure}

\subsection{Magnetic field}
As mentioned before, the bosonized form of the electronic operators in our model is
$\psi_{j,R/L}\sim e^{-irk_{F}x}e^{-i\left(r\phi_{j}-\theta_{j}\right)}$.
The band filling relative to the neutrality point $\nu$ is related to the Fermi momentum as
$k_{F}=\frac{\pi}{2a}\left(1-\nu\right)$. 
The filling $\nu=1$ corresponds to 1 electron per unit-cell, whose length along the wire is 2a, due to the doubled unit-cell introduced by the interwire coupling.
The magnetic field is applied by the ``boost'' transformation $\psi_{j,R/L}\to\psi_{j,R/L}e^{ibjx}$, with $b=edB/\hbar$, and $\Phi_{0}=\frac{h}{e}$ the flux quantum.
Introducing a finite magnetic flux between the wires, the FCI, aFCI, and CDW operators transform as
\begin{align}
    {\cal O}_{{\rm FCI}}^{j}&\sim g_{{\rm FCI}}\left(\psi_{j,R}^{\dagger}\psi_{j,L}\right)^{p}\left(\psi_{j+1,R}^{\dagger}\psi_{j+1,L}\right)^{p}\psi_{j,R}^{\dagger}\psi_{j+1,L}+{\rm h.c.}\nonumber\\
    &=g_{{\rm FCI}}\cos\left[m\left(\phi_{j}+\phi_{j+1}\right)-\theta_{j}+\theta_{j+1}+bx+2mk_{F}x\right], \label{eq:suppOfci}
\end{align}
\begin{align}
    {\cal O}_{{\rm aFCI}}^{j}&\sim g_{{\rm aFCI}}\left(\psi_{j,R}^{\dagger}\psi_{j,L}\right)^{p}\left(\psi_{j+1,R}^{\dagger}\psi_{j+1,L}\right)^{p}\psi_{j+1,R}^{\dagger}\psi_{j,L}+{\rm h.c.}\nonumber\\
    &=g_{{\rm aFCI}}\cos\left[m\left(\phi_{j}+\phi_{j+1}\right)+\theta_{j}-\theta_{j+1}-bx+2mk_{F}x\right], \label{eq:suppOafci}
\end{align}
\begin{align}
    {\cal O}_{{\rm CDW}}^{j}&\sim g_{{\rm CDW}}\left(\psi_{j,R}^{\dagger}\psi_{j,L}\right)^{m}+{\rm h.c.}\nonumber\\
    &=g_{{\rm CDW}}\cos\left(2m\phi+2mk_{F}x\right). \label{eq:suppOcdw}
\end{align}
At filling $\nu_{b=0} = \frac{m-2l}{m}$, with $l$ an integer number, the $2mk_Fx$ factor in all three terms effectively vanishes, and the corresponding phases are commensurate.
(Notice only $l\in\left[-p \, , \, p\right]$ are relevant here, since our model is restricted to $\nu\in\left[0,2\right]$).
With finite magnetic flux per unit cell, $\Phi=2adB$, the commensuration condition for the CDW remains unaltered.
However, for the fractional Chern phases this condition migrates,
\begin{equation}
    \nu^*_{\rm FCI/aFCI} = \nu_{b=0} \pm \frac{1}{m}\frac{\Phi}{\Phi_0}.
\end{equation}
From the well-known Streda formula, 
$\frac{\partial n}{\partial B}= C/\Phi_0$, one confirms the Chern numbers of the FCI and aFCI phases are $1/m$ and $-1/m$, respectively. 

At a given filling factor $\nu$, and magnetic flux $\Phi$, we may define the deviation from commensuration $\delta_\nu = \nu-\nu_{b=0}$.
At finite deviation and/or magnetic fields, the cosines in Eqs.~\eqref{eq:suppOfci}--\eqref{eq:suppOcdw} may oscillate along the direction of the wires.
The spatial period of oscillations depends of course on $\delta \nu$ and $\Phi$.
Within the RG treatment, it is a reasonable approximation~\cite{commIncommRG} to treat this period length as the length scale at which the corresponding cosine is cut off, and the system realizes the incommensurability.
Recalling the short-distance cutoff $\alpha=\alpha_0e^\ell$, we approximate the thresholds at which the different multi-particle terms are cut off as
\begin{equation}
    \ell_{\rm FCI}^{*}=
    -\ln \left[\frac{m}{2}\left(\delta\nu-\frac{1}{m}\frac{\Phi}{\Phi_{0}}\right)\right]\label{eq:ellFCI},
\end{equation}
\begin{equation}
    \ell_{\rm aFCI}^{*}=
    -\ln \left[\frac{m}{2}\left(\delta\nu+\frac{1}{m}\frac{\Phi}{\Phi_{0}}\right)\right]\label{eq:ellaFCI},
\end{equation}
\begin{equation}
    \ell_{\rm CDW}^{*}=
    -\ln \left(\frac{m}{2}\delta\nu\right)\label{eq:ellCDW},
\end{equation}
which were obtained by setting $\alpha_{0}\approx\frac{a}{2\pi}$.

In order to introduce the incommensurability cutoff in a smooth way, we introduce the functions~\cite{commincommoregshavit}
\begin{equation}
    c_i\left(\ell\right) = \frac{1}{\left(e^{\ell-\ell_i^*}\right)^{\gamma}+1},
\end{equation}
where $\gamma$ sets the smoothness of the transition.
At $\ell \gg \ell_i^*$, this function vanishes exponentially fast.
In the opposite limit, $c_i$ tends to unity.
We use $c_i$ in the RG equations to cut off the effect of the incommensurate terms at a finite RG time.
For completeness, the full set of RG equations is given by
\begin{equation}
\begin{aligned}
    \frac{d}{d\ell}y_{{\rm FCI}}&=\left(2-d_{{\rm FCI}}+\frac{m}{2}K_{\rho}K_{\sigma}^{-1}\tilde{V}\right)y_{{\rm FCI}},\\
    \frac{d}{d\ell}y_{{\rm aFCI}}&=\left(2-d_{{\rm FCI}}-\frac{m}{2}K_{\rho}K_{\sigma}^{-1}\tilde{V}\right)y_{{\rm aFCI}},\\
    \frac{d}{d\ell}y_{{\rm CDW}}&=\left(2-d_{{\rm CDW}}\right)y_{{\rm CDW}},\\
    \frac{d}{d\ell}y_{\phi}&=\left(2-2K_{\sigma}\right)y_{\phi},\\
    \frac{d}{d\ell}K_{\rho}^{-1}&=\frac{m^{2}}{2}\left(c_{\rm FCI}\left(\ell\right)y_{{\rm FCI}}^{2}+c_{\rm aFCI}\left(\ell\right)y_{{\rm aFCI}}^{2}+c_{\rm CDW}\left(\ell\right)y_{{\rm CDW}}^{2}\right),\\
    \frac{d}{d\ell}K_{\sigma}&=\frac{1}{2}\left(c_{\rm FCI}\left(\ell\right)y_{{\rm FCI}}^{2}+c_{\rm aFCI}\left(\ell\right)y_{{\rm aFCI}}^{2}\right)-K_{\sigma}^{2}\left(\frac{m^{2}}{2}c_{\rm cdw}\left(\ell\right)y_{{\rm CDW}}^{2}+2y_{\phi}^{2}\right),\\
    \frac{d}{d\ell}\tilde{V}&=m\left(c_{\rm FCI}\left(\ell\right)y_{{\rm FCI}}^{2}-c_{\rm aFCI}\left(\ell\right)y_{{\rm aFCI}}^{2}\right).
\end{aligned}\label{eq:RGwithsmoothcutoff}
\end{equation}

\section{Generalized BKT equations}

When $d_{{\rm FCI}}\approx2$ and all other coupling coefficients remain approximately stationary during the RG flow, the FCI--a-FCI competition is at its strongest.
In that limit, the competition may be compactly described using just the following three flow equations,
\begin{equation}
\begin{aligned}
    \frac{d}{d\ell}y_{{\rm FCI}}=\frac{m}{2}\frac{K_{\rho}}{K_{\sigma}}Vy_{{\rm FCI}},\\
    \frac{d}{d\ell}y_{{\rm aFCI}}=-\frac{m}{2}\frac{K_{\rho}}{K_{\sigma}}Vy_{{\rm aFCI}},\\
    \frac{d}{d\ell}\Tilde{V}=m\left(y_{{\rm FCI}}^{2}-y_{{\rm aFCI}}^{2}\right).
\end{aligned}
\end{equation}
Let us rescale the coefficients to a more recognizable form, 
$y_{1}=m\sqrt{\frac{K_{\rho}}{2K_{\sigma}}}y_{{\rm FCI}}$,
$y_{2}=m\sqrt{\frac{K_{\rho}}{2K_{\sigma}}}y_{{\rm aFCI}}$,
$x=\frac{m}{2}\frac{K_{\rho}}{K_{\sigma}}\Tilde{V}$,
so that we may write,
\begin{equation}
\begin{aligned}
    \frac{d}{d\ell}y_{1}=xy_{1},\\
    \frac{d}{d\ell}y_{2}=-xy_{2},\\
   \frac{d}{d\ell}x=y_{1}^{2}-y_{2}^{2}.\label{eq:genBKTeqns}
\end{aligned}
\end{equation}
Clearly, taking either $y_{1}$ or $y_{2}$ $\to0$ recovers a simple Brezinskii-Kosterlitz-Thouless (BKT) sort of RG flow which is well understood. 
However, the equations above are slightly more complicated.
We begin tackling these equations by identifying two integral of motion,
\[
A=y_{1}y_{2},
\]
\[
B=x^{2}-y_{1}^{2}-y_{2}^{2},
\]
which remain invariant under the RG flow. 
Since we are most interested in the pure FCI--aFCI competition, we focus on the case where the initial value of $\Tilde{V}$ (or $x$) is zero. 
Using the integrals of motion, we find that throughout the RG evolution,
\[
y_{1,0}y_{2,0}=y_{1}y_{2},
\]
\[
-y_{1,0}^{2}-y_{2,0}^{2}=x^{2}-y_{1}^{2}-y_{2}^{2},
\]
where $y_{i,0}$ are the initial values of the coupling constants.
After some straightforward manipulation we obtain
\begin{align*}
\left(xy_{1}\right) & ^{2}=\left(y_{1}^{2}-y_{1,0}^{2}\right)\left(y_{1}^{2}-y_{2,0}^{2}\right).
\end{align*}
Using this relation in the first equation of~\eqref{eq:genBKTeqns}, we have reduced the flow of $y_{1}$ to a single differential equation,
\begin{equation}
    \frac{d}{d\ell}y_{1}=\sqrt{\left(y_{1}^{2}-y_{1,0}^{2}\right)\left(y_{1}^{2}-y_{2,0}^{2}\right)}.
\end{equation}
Recovering the scale $\ell^{\infty}$, where $y_{1}\to\infty$, we find
\begin{equation}
    \ell^{\infty}=\int_{y_{1,0}}^{\infty}dy\frac{1}{\sqrt{\left(y_{1}^{2}-y_{1,0}^{2}\right)\left(y_{1}^{2}-y_{2,0}^{2}\right)}}=\frac{r}{y_{1,0}}{\rm Re}\left[K\left(r^{2}\right)\right].
\end{equation}
where $K\left(m\right)$ is the complete elliptic integral of the first kind with parameter $m=k^{2}$, and we have defined the ratio $r\equiv\frac{y_{1,0}}{y_{2,0}}$.
Notice we are always concerned with the case $r\geq 1$, since the aFCI phase cannot triumph over the FCI.
Having found the RG time at the divergence of $y_{1}$, we may evaluate the energy scale of the gap that opens when $y_{1}$ flows to strong coupling by $\Delta_{{\rm FCI}}=\Lambda_{0}\exp\left(-\ell^{\infty}\right)$, with $\Lambda_{0}$ the initial energy cutoff energy scale. 

In terms of the parameter $r=y_{{\rm FCI},0}/y_{{\rm aFCI},0}$, there exist two particular limits of interest. 
First, if the inhibitory $y_{{\rm aFCI}}$ does not exist (maximally chiral limit), or starts off significantly smaller compared to $y_{{\rm FCI}}$, $r\to\infty$ and
\begin{equation}
    \Delta_{{\rm FCI}} \propto
    \exp \left(-\frac{\pi}{m}\sqrt{\frac{K_\sigma}{2K_\rho}}\frac{1}{ y_{{\rm FCI},0}}\right).
\end{equation}
Notice the dependence on $y_{{\rm FCI},0}$ in the power-law, which is the familiar BKT form. 
In the other interesting limit, $y_{{\rm FCI},0}$ and $y_{{\rm aFCI},0}$ start off at almost the same value, $r\to1$.
Expanding in the deviation of the initial ratio from unity, one finds
\begin{equation}
    \Delta_{{\rm FCI}}\propto
    \left(\frac{y_{{\rm FCI},0}/y_{{\rm aFCI},0}-1}{8}\right)^{\frac{\pi}{m}\sqrt{\frac{K_{\sigma}}{2K_{\rho}}}\frac{1}{y_{{\rm FCI},0}}}.
\end{equation}
This expression possesses a similar power-law behavior as above, yet is \textit{further suppressed} by the small base of the exponent. 
It is instructive to employ the definition $\xi=2d/\log r$ to the last expression, and to obtain (in the appropriate $r\to1$ or $\xi\to\infty$ limit)
\begin{equation}
    \Delta_{{\rm FCI}}\propto
   \left(\frac{d}{4\xi}\right)^{\frac{\sqrt{K_{\sigma}/\left(2K_{\rho}\right)}}{my_{{\rm FCI},0}}}.
\end{equation}
We further emphasize that $\xi$ is intimately connected to the violation of the so-called trace condition far from ideality, see Eq.~\eqref{eq:lgeometricBIG}.
Thus, in this strong competition regime, we have directly shown how the FCI gap is suppressed as a result of ``poor'' quantum geometry.

\begin{figure}
\begin{centering}
\includegraphics[width=10cm]{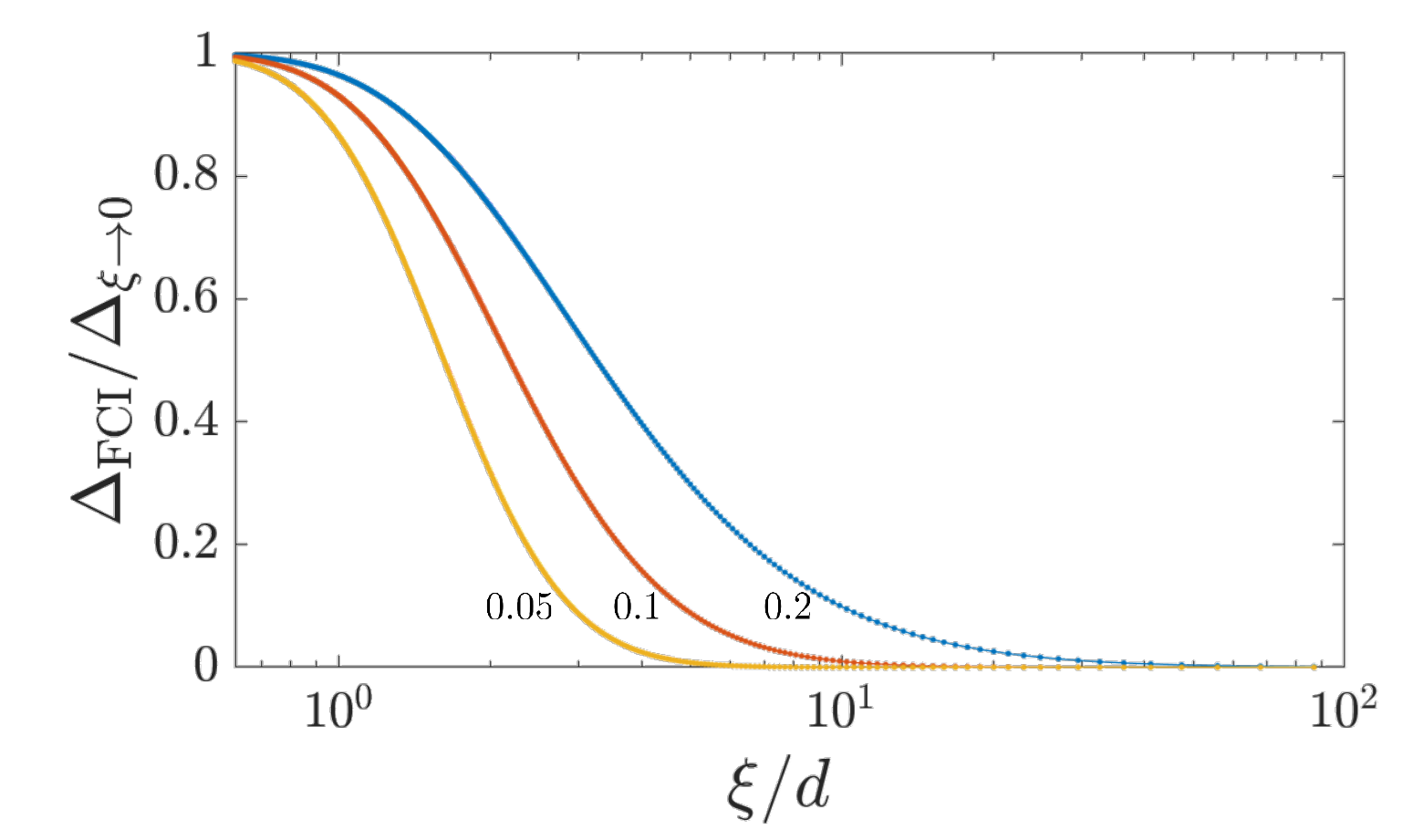}
\par\end{centering}
\caption{The FCI gap calculated in the regime ofinterest above, as a function
of the correlation length $\xi=2d/\log\left(\frac{y_{{\rm FCI},0}}{y_{{\rm aFCI},0}}\right)$.
The three different plots correspond to different initial values of
the paramtere $y_{1,0}=m\sqrt{\frac{K_{\rho}}{2K_{\sigma}}}y_{{\rm FCI},0}$.
As $\xi$ grows and the quantum geometry indicators move further away
from ideality, the size of the correlated FCI gap shrinks substantially.}

\end{figure}

\section{Strong coupling}
Let us write the full Hamiltonian as
\begin{equation}
    H = \int dx \left[ {\cal H}_0 + {\cal H}_{\rm f.s.}
    + {\cal H}_{\rm FCI}+ {\cal H}_{\rm aFCI}+ {\cal H}_{\rm CDW}\right],\label{eq:strongcouplingfull}
\end{equation}
where we express the different terms as
\begin{equation}
    {\cal H}_{0}=\frac{1}{2\pi}\sum_{j}\left[\left(u+V^{0}\right)\left(\partial_{x}\phi_{j}\right)^{2}+\left(u-V^{0}\right)\left(\partial_{x}\theta_{j}\right)^{2}\right],
\end{equation}
\begin{equation}
    {\cal H}_{{\rm f.s.}}=\frac{1}{2\pi}\sum_{j\neq k}\left[\partial_{x}\phi_{j}V_{\phi}^{\left|j-k\right|}\partial_{x}\phi_{k}+\partial_{x}\theta_{j}V_{\theta}^{\left|j-k\right|}\partial_{x}\theta_{k}\right],
\end{equation}
\begin{equation}
    {\cal H}_{{\rm FCI}}=\frac{g_{{\rm FCI}}}{2\pi^{2}}\sum_{j}\cos\left[m\left(\phi_{j}+\phi_{j+1}\right)+\theta_{j}-\theta_{j+1}\right],
\end{equation}
\begin{equation}
    {\cal H}_{{\rm aFCI}}=\frac{g_{{\rm aFCI}}}{2\pi^{2}}\sum_{j}\cos\left[m\left(\phi_{j}+\phi_{j+1}\right)-\theta_{j}+\theta_{j+1}\right],
\end{equation}
\begin{equation}
    {\cal H}_{{\rm CDW}}=\frac{g_{{\rm CDW}}}{2\pi^{2}}\sum_{j}\cos\left(2m\phi_{j}\right).
\end{equation}
As usual, the fields $\phi_j,\theta_j$ correspond to the bosonized fields on the $j$ wire. 
In the above we have assumed translation invariance, as well as conservation of time-reversal symmetry by the forward-scattering part of the interaction.

It is instructive to make an intermediate step, and define the following chiral operators,
\begin{equation}
    \varphi_{j}^{R/L}=\frac{\theta_{j}}{m}\pm\phi_{j},
\end{equation}
which obey the commutation relation 
\begin{equation}
\left[\varphi_{i}^{r}\left(x\right),\varphi_{j}^{r'}\left(x'\right)\right]=\frac{i\pi}{m}r\delta_{rr'}\delta_{ij}{\rm sgn}\left(x-x'\right).\label{eq:chiralcommutationrelations}
\end{equation}
In terms of these chiral operators, the Hamiltonian is
\begin{align}
        {\cal H}_{0}+{\cal H}_{{\rm f.s.}}&=
        \frac{mv}{4\pi}\sum_{j}\left[\left(\partial_{x}\varphi_{j}^{R}\right)^{2}+\left(\partial_{x}\varphi_{j}^{L}\right)^{2}\right]-\frac{\tilde{V^{0}}}{4\pi}\sum_{j}\partial_{x}\varphi_{j}^{R}\partial_{x}\varphi_{j}^{L}
        \nonumber\\
        &+\frac{1}{4\pi}\sum_{j\neq k}\partial_{x}\varphi_{j}^{r}V_{rr'}^{\left|j-k\right|}\partial_{x}\varphi_{k}^{r'}
\end{align}
\begin{equation}
    {\cal H}_{{\rm FCI}}=\frac{g_{{\rm FCI}}}{2\pi^{2}}\sum_{j}\cos\left[m\left(\varphi_{j}^{R}-\varphi_{j+1}^{L}\right)\right]
\end{equation}
\begin{equation}
    {\cal H}_{{\rm aFCI}}=\frac{g_{{\rm aFCI}}}{2\pi^{2}}\sum_{j}\cos\left[m\left(\varphi_{j}^{L}-\varphi_{j+1}^{R}\right)\right],
\end{equation}
\begin{equation}
    {\cal H}_{{\rm CDW}}=\frac{g_{{\rm CDW}}}{2\pi^{2}}\sum_{j}\cos\left[m\left(\varphi_{j}^{L}-\varphi_{j}^{R}\right)\right],
\end{equation}
with the re-defined constants $v=\frac{\left(1+m^{2}\right)u+\left(1-m^{2}\right)V^{0}}{2m}$, $\tilde{V^{0}}=\left(1-m^{2}\right)u+\left(1+m^{2}\right)V^{0}$, $V_{rr'}^{\left|i-j\right|}=\frac{V_{\phi}^{\left|i-j\right|}}{2}\left(2\delta_{rr'}-1\right)+\frac{m^{2}V_{\theta}^{\left|i-j\right|}}{2}$.
Taken together with the chiral operators commutation relations, Eq.~\eqref{eq:chiralcommutationrelations}, we may interpret the Hamiltonian in a different way. 
Each wire has been effectively transformed into a narrow fractional quantum Hall strip analogous to filling $\nu=1/m$, whose chiral edge states have the velocity v. 
The constants $\tilde{V^{0}}$ and $V_{rr'}^{\left|i-j\right|}$ determine a forward-scattering interaction Hamiltonian operating between these chiral edge states throughout the system.
The multiparticle backscattering terms now couple neighboring edge states with $m$-particle processes.
Once more, unlike the fractional quantum hall case, the coupling is not entirely chiral: ${\cal H}_{\rm FCI}$ competes with ${\cal H}_{\rm aFCI}$ and ${\cal H}_{\rm CDW}$.
Due to this competition, a gapped phase which is not compatible the $\nu=1/m$ fractional quantum hall effect may form.

We note that one may define the following quasiparticle operators~\cite{wireconstructionKanePrl,wireconstructionkaneteoPrb},
$\Psi_{{\rm QP},j}^{R/L} \sim e^{i\varphi_j^{R/L}}$,
which are not physical operators by themselves (they cannot be built out of the local electron operators).
However, in the gapped FCI phase, it can be shown these quasiparticles possess fractional abelian statistics by constructing local operators that transfer quasiparticles through the system~\cite{wireconstructionkaneteoPrb}.

Finally, one may construct the fermionic operators
$\Psi_{j}^{R/L} \sim e^{im\varphi_j^{R/L}}$,
in terms of which the cosine terms in the Hamiltonian are tunneling processes of fermions between the edge-states in different quantum hall strips.
The fact that these are indeed fermionic operators may be easily understood by considering the commutation relations, 
\begin{equation}
    \left[m\varphi_{i}^{r}\left(x\right),m\varphi_{j}^{r'}\left(x'\right)\right]=i\pi mr\delta_{rr'}\delta_{ij}{\rm sgn}\left(x-x'\right)=\left(2n+1\right)i\pi r\delta_{rr'}\delta_{ij}{\rm sgn}\left(x-x'\right),
\end{equation}
which differ only by an integer multiple of $2\pi$ from the commutation relations of the original chiral bosonic operators in terms of which the bosonization of the bare electronic Hamiltonian was performed. 
As opposed to the bare electronic operators, which have a scaling dimension of $1/2$, these fermionic operators have an enlarged scaling dimension of $m/2$, a characteristic of the chiral Luttinger liquid at the fractional quantum Hall edges.

Let us now consider the strong coupling limit of the Hamiltonian~\eqref{eq:strongcouplingfull}, where some (or all) of the multiparticle terms dominate all other energy scales in the problem.
Denoting the strong-coupling value of $g_i/\left(\pi\right)$ as $G_i$, we write the Hamiltonian density as
\begin{equation}
    {\cal H}_{\rm strong} = 
    \sum_j
    \left[
    i\tilde{v}\left({\Psi}_{j,R}^{\dagger}\partial_{x}{\Psi}_{j,R}-{\Psi}_{j,L}^{\dagger}\partial_{x}{\Psi}_{j,L}\right)+G_{{\rm FCI}}{\Psi}_{j,R}^{\dagger}{\Psi}_{j+1,L}+G_{{\rm aFCI}}{\Psi}_{j,L}^{\dagger}{\Psi}_{j+1,R}+G_{{\rm CDW}}{\Psi}_{j,R}^{\dagger}{\Psi}_{j,L}+{\rm h.c.}
    \right]+\dots,
\end{equation}
where we have included a linear dispersion along the wires for these chiral fermions with some renormalized velocity $\tilde{v}$ for concreteness.
The $\dots$ represent subdominant interaction terms which cannot open a spectral gap.
${\cal H}_{\rm strong}$ is readily diagonalized, with the spectrum
\begin{equation}
    E_{\rm strong}=\pm\sqrt{\left(\tilde{v}k_{x}\right)^{2}+\left[G_{{\rm CDW}}+\left(G_{{\rm FCI}}+G_{{\rm aFCI}}\right)\cos k_{y}\right]^{2}+\left(G_{{\rm FCI}}-G_{{\rm aFCI}}\right)^{2}\sin^{2}k_{y}}.
\end{equation}
Here, $k_{x}$ ($k_{y}$) the momentum in the longitudinal (transverse) direction. 
In the regime where the CDW is subdominant to the FCI phases, $G_{{\rm CDW}}\leq G_{{\rm FCI}}+G_{{\rm aFCI}}$, the spectral gap is
\begin{equation}
    \Delta E_{\rm strong}=2\left|G_{{\rm FCI}}-G_{{\rm aFCI}}\right|\sqrt{1-\left(\frac{G_{{\rm CDW}}}{G_{{\rm FCI}}+G_{{\rm aFCI}}}\right)^{2}}.
\end{equation}
If we parameterize in a similar way to the discussion in Sec.~\ref{appsec:yfz},
$2G_{\rm F}^2=G_{\rm FCI}^2 + G_{\rm aFCI}^2$, $2G_{\rm F}^2z=G_{\rm FCI}^2 - G_{\rm aFCI}^2$, and expand away from optimal quantum geometry ($z\ ll 1$), we may re-write the gap expression as
\begin{equation}
    \Delta E_{\rm strong} = z\sqrt{G_{\rm F}^2-G_{\rm CDW}^2} +O\left(z^3\right).
\end{equation}

The strong coupling expression reveals that the many-body gap relates directly to the competition between the FCI and the disruptive aFCI phase, with the gap vanishing linearly in their difference.
We emphasize again that the relative strength of the anomalous $G_{\rm aFCI}$ (or the magnitude of $z$) is related to the quantum geometry of the parent Chern band.
Thus, we establish the connection between quantum geometry and the stabilization of the FCI phase in our model even in the strong coupling limit.

\section{Further implications of the coupled wires construction}

\subsection{FCI promotion by periodic modulation}
Consider a periodic modulation of the density, such that the density in even wires is $\nu_{\rm frac.}+\delta\nu_{\rm mod.}$, whereas on odd wires it is $\nu_{\rm frac.}-\delta\nu_{\rm mod.}$.
The CDW part of the Hamiltonian will now read
\begin{equation}
    {\cal H}_{\rm CDW} = \frac{g_{\rm CDW}}{2\pi^2}
    \cos\left(m\sqrt{2}\phi_{\rho}\right)
    \cos\left(m\sqrt{2}\phi_{\sigma}+m\frac{\pi}{a}\delta\nu_{\rm mod.} x\right)
    +\frac{g_{\phi}}{2\pi^{2}}\cos\left(\sqrt{8}\phi_{\sigma}+2\frac{\pi}{a}\delta\nu_{\rm mod.} x\right).\label{eq:modifiedcdwH}
\end{equation}
Similar to our modifications leading to the RG flow in Eq.~\eqref{eq:RGwithsmoothcutoff}, the flow associated with terms quadratic in $g_{\rm CDW}$ and $g_\phi$ now acquire the respective multiplicative constants
\begin{equation}
    c_{\rm CDW}\left(\ell\right) = \left(1+e^{\gamma\left(\ell+\frac{m}{2}\log\delta\nu_{\rm mod.}\right)}\right)^{-1},
\end{equation}
\begin{equation}
    c_{\phi}\left(\ell\right) = \left(1+e^{\gamma\left(\ell+\log\delta\nu_{\rm mod.}\right)}\right)^{-1}.
\end{equation}

\begin{figure}
\begin{centering}
\includegraphics[width=14cm]{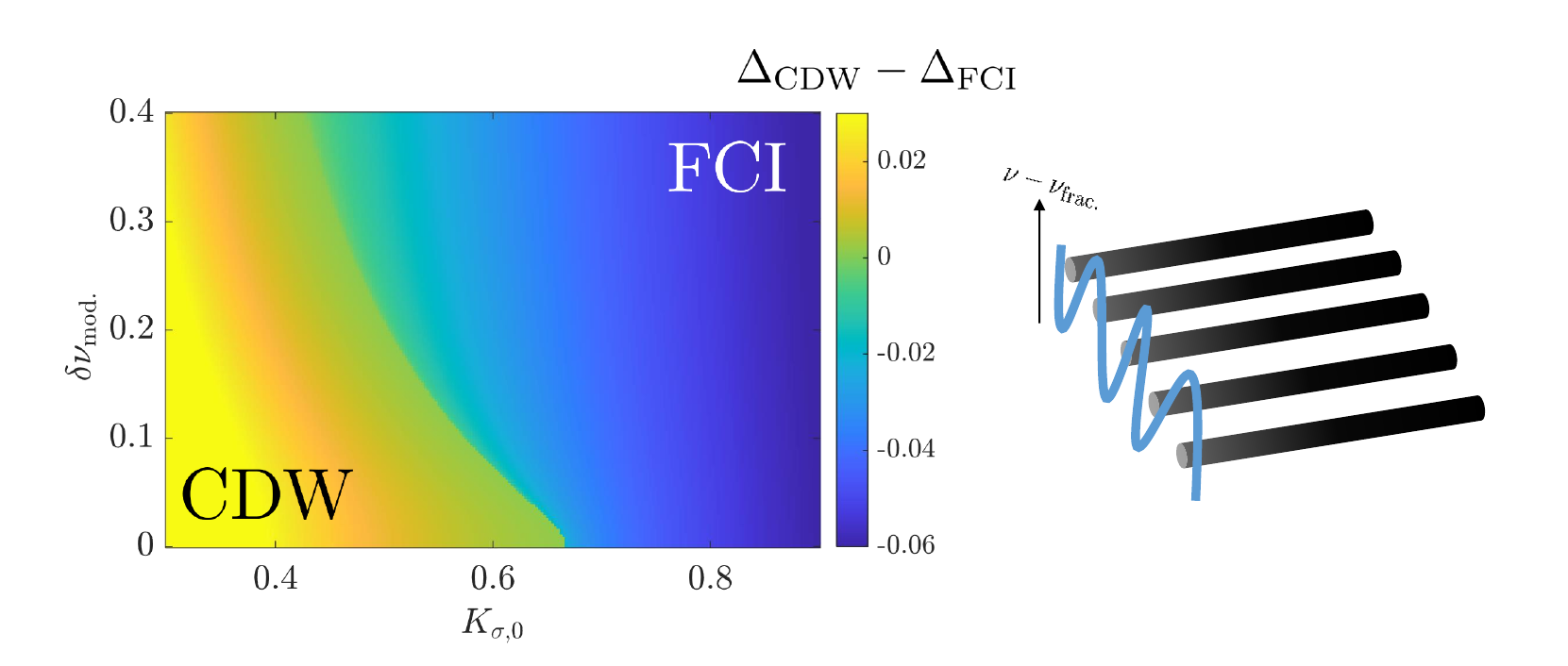}
\par\end{centering}
\caption{Left:
    Phase diagram obtained by the RG flow, with CDW dephasing parameter $\delta\nu_{\rm mod.}$, which modifies the cutoff of the CDW Hamiltonian terms [Eq.~\eqref{eq:modifiedcdwH}].
    We plot the difference between the gap proxies $\Delta_{\rm CDW}-\Delta_{\rm FCI}$ with the parameters
    $m=3$ and initial conditions $z=0.2$, $K_{\rho,0}=1/25$, $y_{F,0}=0.03$, $y_{\rm CDW,0}=0.08$, $y_{\phi,0}=0.1$, $\tilde{V}_0=0$.
    Right:
    Schematic illustration of the proposed modulation along the array of wires.
    }\label{fig:CDWdephasing}
\end{figure}

In Fig.~\ref{fig:CDWdephasing} we demonstrate the effect of the density modulation on the phase diagram.
Namely, the modulation leads to dephasing and destabilization of the CDW phase at shorter and shorter time scales.
In turn, this leads to promotion of the FCI, and its stabilization over larger areas of parameter space.
Thus, our coupled wires model points at some interesting opportunities in lattice and band engineering, if one aims to acheive a robust FCI phase.

\subsection{CDW stabilized by a magnetic field}
Recent experiments in moir\'e graphene heterostructures~~\cite{magneticWignerCrystal,magneticWignerCrystal2} have observed a peculiar trend, where a CDW or Wigner crystal phase is stabilized at fractional band filling by applying a perpendicular magnetic field.
Surprisingly, a similar effect may be observed within our model at a certain parameter regime.

Fixing the density at $\nu_{\rm frac.}$, in the presence of finite magnetic flux the RG-time thresholds are
\begin{equation}
    \ell^*_{\rm FCI}=\ell^*_{\rm aFCI}=-\ln \left|\frac{\Phi}{2\Phi_0}\right|,
\end{equation}
wheres there is no threshold for the CDW term.
Application of a magnetic field at the appropriate fractional density thus renders the FCI phases incommensurate, effectively cutting them off at shorter length scales, which may lead to CDW stabilization.
In a sense, it is the analogous effect to the one described in the previous section -- now the magnetic field ``dephases'' the FCI and aFCI, potentially promoting the CDW.

As illustrated in Fig.~\ref{fig:CDW_stabilized}, application of a magnetic field stabilizes the CDW at the expanse of the FCI phase.
The CDW gap itself gradually increases with magnetic field, suggesting that our proposed model may help identify the cause for magnetic-field-induced stabilization of Wigner crystals at fractional filling.

\begin{figure}
\begin{centering}
\includegraphics[width=14cm]{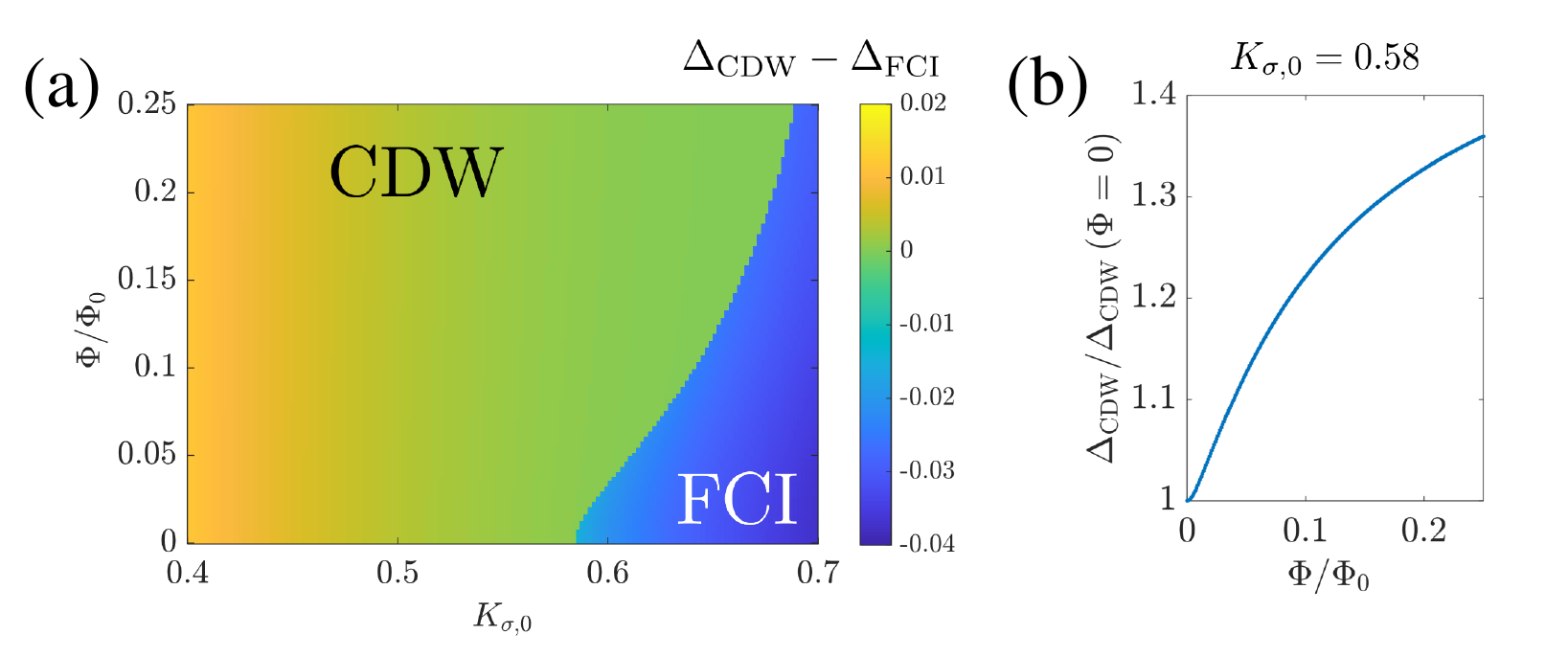}
\par\end{centering}
\caption{(a):
    Phase diagram obtained by the RG flow, in the presence of a finite magnetic field.
    We plot the difference between the gap proxies $\Delta_{\rm CDW}-\Delta_{\rm FCI}$ with the parameters
    $m=3$ and initial conditions $z=0.05$, $K_{\rho,0}=1/15$, $y_{F,0}=0.03$, $y_{\rm CDW,0}=0.08$, $y_{\phi,0}=0.06$, $\tilde{V}_0=0$.
    (b)
    A vertical of panel (b) taken at a specific initial value of $K_{\sigma,0}$ (indicated above the panel).
    The CDW is stabilized with increased magnetic field.
    }\label{fig:CDW_stabilized}
\end{figure}
\end{widetext}

\bibliographystyle{apsrev4-2}
\end{document}